\documentclass[11pt]{article}

\usepackage[final]{acl}

\usepackage{amsmath}
\usepackage{amsfonts}
\usepackage{amssymb}
\usepackage{amsthm}

\usepackage{times}
\usepackage{latexsym}
\usepackage[T1]{fontenc}
\usepackage{microtype}
\usepackage{inconsolata}

\usepackage{booktabs}
\usepackage{multirow}
\usepackage{makecell}
\usepackage{tabularx}
\usepackage{graphicx}
\usepackage{subcaption}

\usepackage{enumitem}

\usepackage[dvipsnames,table]{xcolor}

\usepackage[utf8]{inputenc}
\usepackage{kotex}

\usepackage{hyperref}

\newcommand{\same}[1]{\textcolor{blue}{#1}}
\newcommand{\diff}[1]{\textcolor{red}{#1}}

\definecolor{lightgray}{gray}{0.9}

\definecolor{lightgray}{gray}{0.9}

\newcommand{\dec}[1]{\textcolor{red}{(\ensuremath{\downarrow}~#1)}}

\title{Towards Privacy-Preserving Large Language Model: Text-free Inference Through Alignment and Adaptation}

\author{
  \textbf{Jeongho Yoon}$^{1}$,
  Chanhee Park$^{1}$,
  Yongchan Chun$^{1}$,
  Hyeonseok Moon$^{2}$,
  Heuiseok Lim$^{1}$\thanks{Corresponding author.} \\
  $^{1}$Department of Computer Science and Engineering, Korea University \\
  $^{2}$Samsung Mobile eXperience Business \\
  \texttt{\{aa007878,pch7678,cyc9805,limhseok\}@korea.ac.kr} \\
  \texttt{hyns.moon@samsung.com}
}

\begin{document}
\maketitle
\begin{abstract}

Current LLM-based services typically require users to submit raw text regardless of its sensitivity. While intuitive, such practice introduces substantial privacy risks, as unauthorized access may expose personal, medical, or legal information. Although prior defenses strived to mitigate these risks, they often incur substantial computational overhead and degrade model performance.
To overcome this privacy--efficiency trade-off, we introduce \textbf{Privacy-Preserving Fine-Tuning (PPFT)}, a novel training pipeline that eliminates the need for transmitting raw prompt text while maintaining a favorable balance between privacy preservation and model utility for both clients and service providers. 
Our approach operates in two stages: first, we train a client-side encoder together with a server-side projection module and LLM, enabling the server to condition on $k$-pooled prompt embeddings instead of raw text; second, we fine-tune the projection module and LLM on private, domain-specific data using noise-injected embeddings, allowing effective adaptation without exposing plain text prompts and requiring access to the decoder’s internal parameters.
Extensive experiments on domain-specific and general benchmarks demonstrate that PPFT achieves a striking balance between privacy and utility, maintaining competitive performance with minimal degradation compared to noise-free upper bounds. 

\end{abstract}

\section{Introduction}
\begin{figure}[h]
\centering
 \includegraphics[width=\columnwidth]{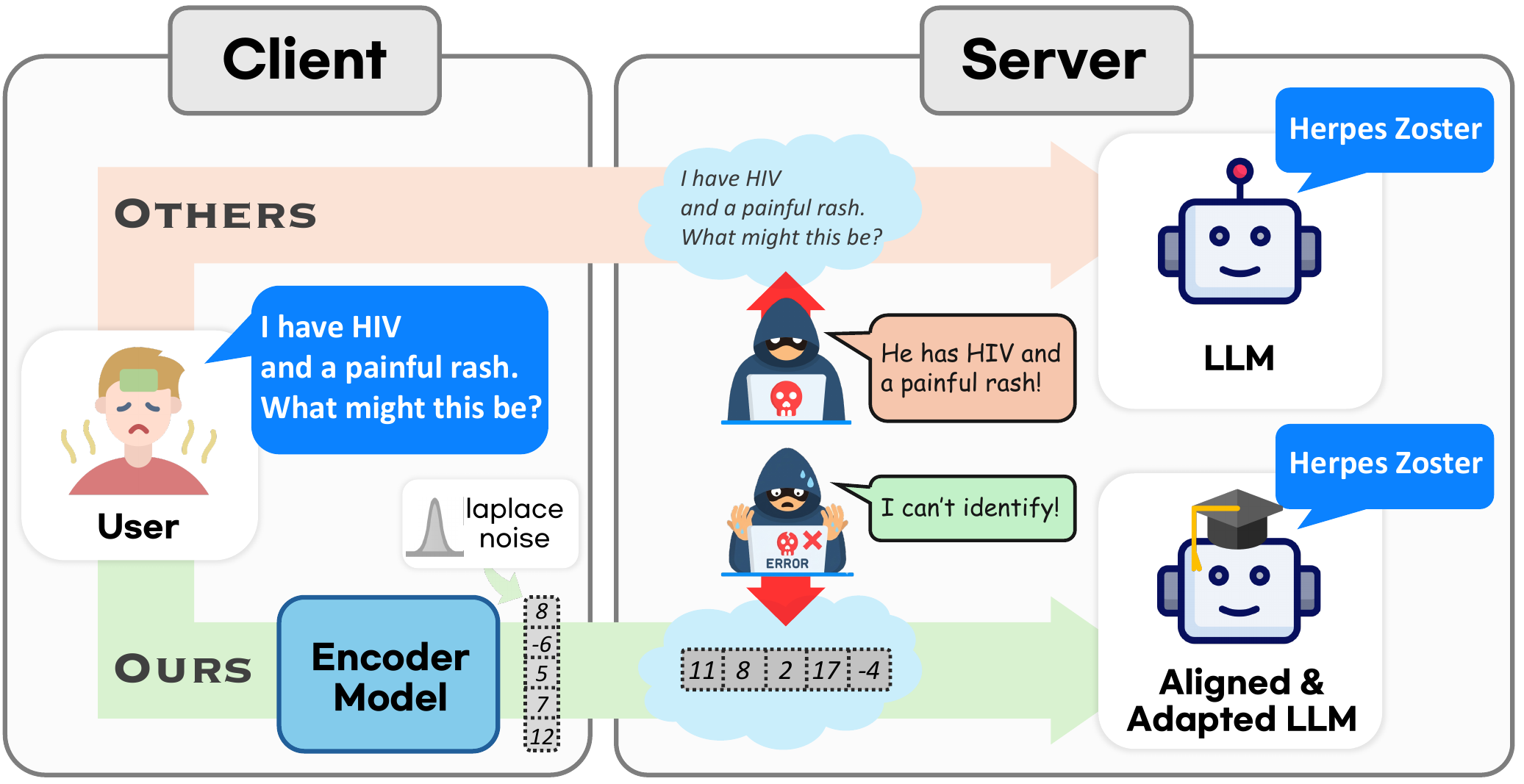} 
\caption{While conventional services expose plain text prompts to the server, PPFT transmits only obfuscated embeddings to prevent prompt inference and mitigate privacy risks.}
\label{fig:Prev_llm}
\end{figure}

Driven by rapid advances, large language models (LLMs) now serve as effective tools across a wide range of domains that require specialized expertise, including healthcare, law, and finance~\cite{wiggins2022opportunities, achiam2023gpt, singhal2025toward, guha2023legalbench}. 
Several studies have actively explored their capabilities in professional clinical assistance in healthcare~\cite{singhal2025toward}, as well as in legal reasoning~\cite{guha2023legalbench, huang2023lawyer}.

In practical use-cases, LLMs are typically deployed in cloud-based MLaaS (Machine Learning as a Service) settings that require transmitting prompts as \emph{plain text} \cite{comanici2025gemini, achiam2023gpt}. However, once the original prompt is sent in plain text, we argue that the natural-language input becomes vulnerable to adversarial interception during transmission and to unauthorized access in the event of a cloud infrastructure breach, creating a fundamental privacy vulnerability~\cite{chong2024casper, carlini2021extracting}. Processing sensitive content such as medical or legal records in this written form not only risks immediate leakage via eavesdropping or insider misuse, but can also lead to persistent exposure through system logs and downstream training pipelines, constituting a critical security hazard~\cite{kibriya2024privacy}.

To mitigate privacy risks, prior work explored transmitting embeddings instead of raw text~\cite{mai2023split}. However, recent findings demonstrate that even heuristically noised embeddings remain vulnerable to generative inversion attacks that reconstruct semantically faithful text~\cite{morris2023text, li2023sentence}. This highlights a critical flaw: embedding transmission, even with ad hoc noise, lacks strong privacy guarantees.
Meanwhile, cryptographic protocols and existing training-stage defenses often incur prohibitive costs or remain fragile against reconstruction, limiting their scalability~\cite{hao2022iron, lin2024inversion}.
Consequently, a unified framework that eliminates prompt text transmission during both inference and fine-tuning while preserving efficiency and performance remains underexplored.

To address this gap, we propose \textbf{PPFT (Privacy-Preserving Fine-Tuning)}, which operationalizes the principle of \emph{never sending the prompt} under realistic system constraints. A lightweight client-side encoder first maps the prompt to token-level embeddings, after which PPFT applies $k$-Pooling to aggregate representations over fixed-size token groups, thereby reducing recoverable token-level detail and increasing the difficulty of prompt reconstruction.
To further suppress residual leakage, PPFT injects Laplace noise and transmits only the resulting obfuscated embeddings to the server. The server-side LLM is trained to directly consume these obfuscated embeddings, enabling semantic conditioning without access to prompt text.

Crucially, PPFT enforces the same interface during both inference and fine-tuning, ensuring that raw prompts are never exposed to the server and allowing domain adaptation to proceed without requiring disclosure of the decoder’s internal parameters.

Across medical and legal question answering tasks as well as general-purpose benchmarks, PPFT preserves task performance while exhibiting strong robustness against inversion attacks, achieving practical privacy protection. The main contributions of this paper are as follows:
\begin{itemize}
\item \textbf{Text-free Prompt Interface for Fine-tuning and Inference:}
We propose an end-to-end privacy-preserving pipeline that eliminates prompt text transmission during both inference and fine-tuning via client-side embedding, $k$-Pooling–based compression, and obfuscated embedding transfer.
\item \textbf{Domain-specific Adaptation without Prompt and Model Exposure:}  
We show that effective domain adaptation in sensitive domains is possible without server-side access to raw prompt text and disclosure of proprietary decoder parameters, enabling privacy-preserving fine-tuning under realistic service deployment constraints.
\item \textbf{Inversion-Resistant Obfuscated Embedding Interface:}
We inject Laplace noise into pooled embeddings and train the decoder to operate on obfuscated embedding, improving robustness against prompt reconstruction attacks.
\end{itemize}

\section{Related Work}

\subsection{Prompt Privacy in Cloud-based LLM Services}
\label{sec:prompt_privacy}

Cloud-hosted LLMs are commonly offered as MLaaS via web or API interfaces, where users must transmit prompts to remote servers.
A widely deployed defense is prompt sanitization, which detects and redacts sensitive spans on-device before sending the request~\cite{shen2024fire}.
However, sanitization can miss contextual or implicit disclosures~\cite{ngong2025protecting} and still retains the text-based interface in which the server receives a textual prompt~\cite{chong2024casper}.
Cryptographic inference can hide inputs during computation, but its compute/communication overhead remains prohibitive for large Transformer models in real-time settings~\cite{gilad2016cryptonets, hao2022iron}.

Representation-level alternatives improve efficiency by perturbing embeddings or intermediate states~\cite{feyisetan2020privacy, mai2023split, du2023dp}, but differ substantially in system assumptions and privacy scope. 
DP-Forward~\cite{du2023dp} injects differential privacy noise into the forward computation for fine-tuning and inference, while Split-and-Denoise~\cite{mai2023split} protects inference by executing the embedding layer on the client and applying local DP before server-side processing. 
SentineLLMs~\cite{mishra2024sentinellms} studies secure adaptation with protected inputs, and recent cloud--edge systems such as PRISM~\cite{zhan2026prism} further combine privacy-aware routing with collaborative sketch/refinement execution.
However, these approaches generally focus on inference-time protection, encrypted/secure execution, or adaptive routing, rather than enforcing a single reusable text-free interface under which the server can both perform inference and adapt to private-domain data without observing raw prompts.
Considering these, we define a text-free interface for both inference and fine-tuning: the client transmits only embedding vectors from a client-side encoder, and the server consumes them via a projection-based connection to a high-capacity decoder.

\begin{figure*}[ht]
\centering
 \includegraphics[width=\textwidth]{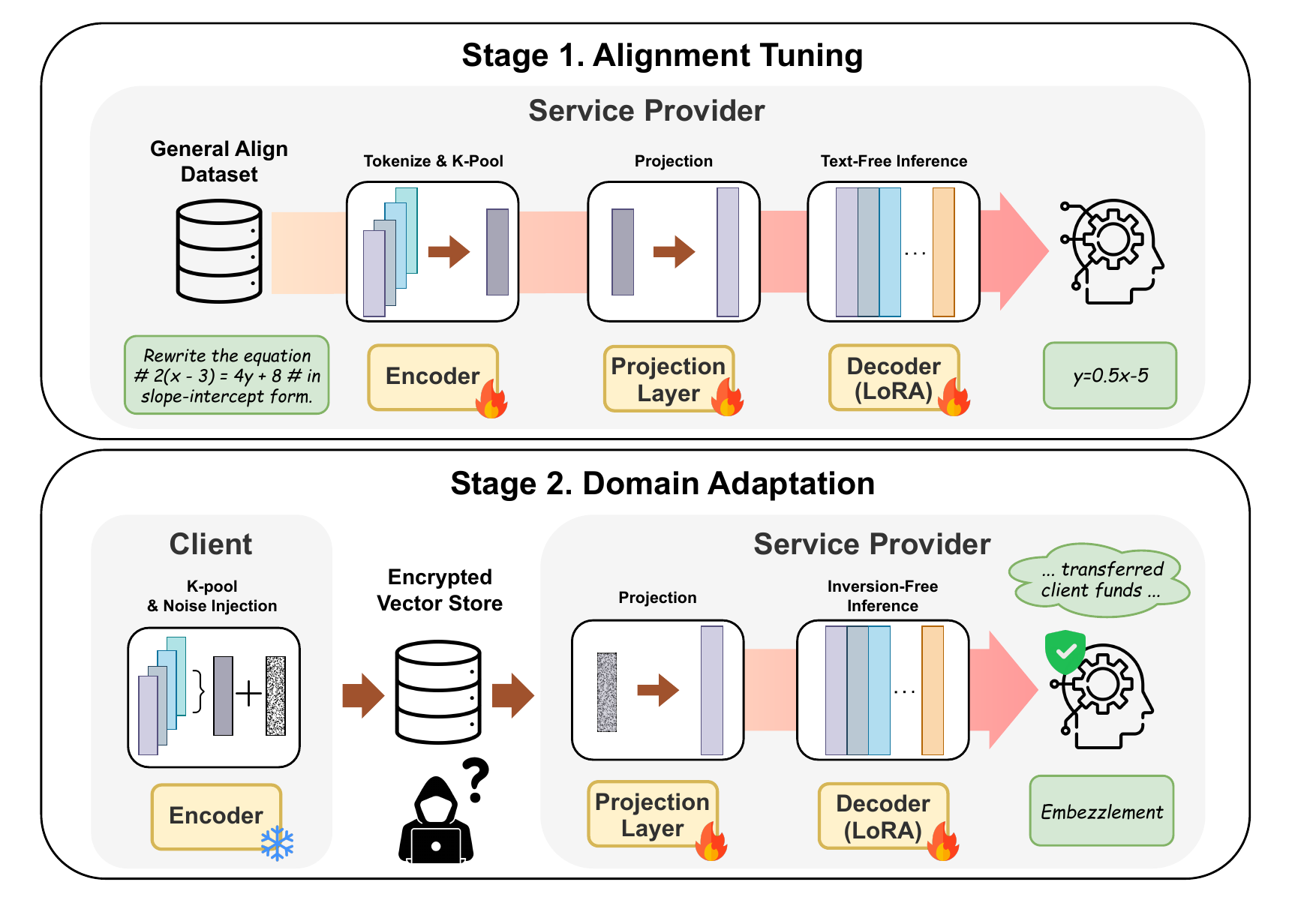} 
\caption{Overview of PPFT. Stage~1 aligns pooled client-side embeddings with the decoder to enable text-free inference. Stage~2 performs domain adaptation using noise-injected embeddings to improve robustness against reconstruction.}
\label{fig:main_fig}
\end{figure*}

\subsection{Embedding Leakage and Inversion Attacks}
\label{sec:inversion}

Although existing studies explore transferring embeddings instead of raw text, it is inherently unsafe: modern text embeddings preserve substantial semantic and contextual information, enabling generative inversion that reconstructs meaningful approximations of the original prompt~\cite{morris2023text, li2023sentence}.
Even when embeddings are obfuscated, dedicated attacks can recover the original input from transformed vectors, underscoring that embedding-only transmission does not guarantee privacy~\cite{zhou2023textobfuscator, lin2024inversion}.
These studies suggest that we can attain an effective protection with noise mechanisms considering reconstructability and decoders trained to operate on noisy inputs.
PPFT instantiates this by $k$-Pooling, noise injection, and decoder training on obfuscated continuous embeddings.

\subsection{Privacy-Preserving Training Beyond Parameter Privacy}
\label{sec:pp_finetuning}

Prior work on privacy-preserving fine-tuning largely targets parameter privacy, aiming to prevent memorization of training data and mitigate membership inference or extraction.
DP-SGD is the canonical approach~\cite{abadi2016deep}, and recent extensions combine DP with PEFT (e.g., LoRA/adapters) to reduce computational and privacy overhead by restricting differentially private updates to a small set of lightweight modules~\cite{yu2021differentially, liu2025differentially}.
However, these methods typically assume the server still receives and processes plain text training prompts, leaving input confidentiality unresolved in MLaaS settings.
Related paradigms such as split learning or federated learning keep raw data local but can leak through intermediate representations or gradients, often requiring additional protections~\cite{qiu2023evaluating}.

Among split-learning-based approaches, Split-and-Privatize~\cite{shen2023split} is particularly related in that it mitigates privacy risks in MaaS fine-tuning by adapting split execution.
However, its primary focus is training-time privacy under split learning, whereas PPFT establishes a reusable embedding-only interface that is consistently maintained across both inference and domain adaptation, with the additional goal of reducing inversion risk through pooling and noise injection.

To address these limitations, we design a text-free interface that protects prompt privacy while keeping the server model opaque to clients: all fine-tuning and inference are carried out using client-produced obfuscated embeddings, allowing adaptation without revealing raw prompts or the server’s decoder parameters.

\section{PPFT}

In this paper, we propose \textbf{Privacy-Preserving Fine-Tuning (PPFT)}, a novel framework that eliminates plain text prompt transmission in MLaaS. As illustrated in Figure~\ref{fig:main_fig}, our approach consists of two stages: (1) alignment of encoder-decoder representations via continuous embeddings, and (2) privacy-preserving domain adaptation with noise injection, enabling a completely text-free inference pipeline.

\subsection{Problem Statement and Notation}
\label{sec:problem_statement}

We aim to construct a text-free prompt interface where the server generates responses conditioned solely on embeddings transmitted from the client, without accessing raw prompt text. Let $\mathbf{x}=(x_1,\dots,x_n)$ be the user prompt and $\mathbf{y}=(y_1,\dots,y_T)$ be the target response.
We utilize a client-side encoder $E_{\phi}$ that outputs hidden representations $\mathbf{H}=E_{\phi}(\mathbf{x})\in\mathbb{R}^{n\times d_e}$, where $\mathbf{H}=[\mathbf{h}_1;\dots;\mathbf{h}_n]$.
The server hosts a causal LLM decoder $D_{\theta}$ which generates $\mathbf{y}$ given a continuous prefix.
To bridge the dimension mismatch between the encoder ($d_e$) and decoder ($d_d$), a trainable projection layer $P_{\psi}$ is employed.

\subsection{Stage 1: Encoder--Decoder Alignment}
\label{sec:stage1}

The objective of Stage 1 is to align the latent spaces of the independent encoder and decoder, enabling the decoder to perform semantic conditioning based on embeddings rather than discrete tokens. This stage establishes the foundation for text-free interaction through token compression and projection.

\paragraph{$k$-Pooling for Token Compression.}
To reduce recoverable token-level detail and increase reconstruction difficulty, we apply block-wise mean pooling to the encoder output $\mathbf{H}$. The pooling function $\mathrm{Pool}_k:\mathbb{R}^{n\times d_e}\to\mathbb{R}^{m\times d_e}$ reduces the sequence length to $m=\lceil n/k\rceil$. The $j$-th pooled vector $\mathbf{u}_j$ is computed as:
\begin{equation}
\mathbf{u}_j=\frac{1}{|I_j|}\sum_{i\in I_j}\mathbf{h}_i,
\label{eq:kpool}
\end{equation}
where $I_j=\{(j-1)k+1,\dots,\min(jk,n)\}$ denotes the index set of tokens in the $j$-th block. The results in the pooled embeddings $\mathbf{U}=[\mathbf{u}_1;\dots;\mathbf{u}_m]$.

\paragraph{Continuous Prefix Injection.}
The pooled embeddings $\mathbf{U}$ are then mapped to the decoder's input space via the projection layer $P_{\psi}$, yielding $\mathbf{Z} = P_{\psi}(\mathbf{U}) \in \mathbb{R}^{m\times d_d}$.
These projected vectors form a continuous conditioning context for the decoder, which directly conditions generation on $\mathbf{Z}$ without any discrete prompt tokens.
The model is trained to minimize the negative log-likelihood of the target sequence $\mathbf{y}$ given the prefix $\mathbf{Z}$:
\begin{equation*}
\mathcal{L}_{\mathrm{align}}(\phi,\psi,\theta) = -\sum_{t=1}^{T}\log p_{\theta}(y_t \mid y_{<t}, \mathbf{Z}).
\label{eq:align_loss}
\end{equation*}
In this stage, we jointly update the encoder $E_{\phi}$, projection layer $P_{\psi}$, and LoRA~\cite{hu2022lora}-adapted decoder $D_{\theta}$ parameters to ensure robust semantic transfer.

\subsection{Stage 2: Text-free Domain Adaptation}
\label{sec:stage2}

Stage 2 focuses on adapting the model to specific domains (e.g., medical, legal) while enforcing strict privacy guarantees. This is achieved by injecting privacy-preserving noise into the embeddings and fine-tuning the server-side components without exposure to raw text.

\paragraph{Noise Injection Mechanism.}
Building upon $\mathbf{U}$ in Eq.~\ref{eq:kpool}, we inject calibrated noise with an interpretation under $d_\chi$-privacy~\cite{feyisetan2020privacy}.
For each row vector in $\mathbf{U}$, we add isotropic Laplace noise, constructed by sampling a direction uniformly from the unit sphere and a magnitude from a Gamma distribution (shape $d_e$, rate $\epsilon$).
We then apply $L_2$ re-normalization as a post-processing step, obtaining $\tilde{\mathbf{U}}$, which we refer to as \emph{obfuscated embeddings}.

\paragraph{Privacy-Preserving Fine-Tuning.}
The server receives only the obfuscated embeddings $\tilde{\mathbf{U}}$ and the target labels $\mathbf{y}$.
It projects $\tilde{\mathbf{U}}$ to $\tilde{\mathbf{Z}} = P_{\psi}(\tilde{\mathbf{U}})$ and fine-tunes the model conditioned on $\tilde{\mathbf{Z}}$.
The client-side encoder $E_{\phi}$ is not fine-tuned in this stage. The optimization target is:
\begin{equation*}
\mathcal{L}_{\mathrm{priv}}(\psi,\theta) = -\sum_{t=1}^{T}\log p_{\theta}(y_t \mid y_{<t}, \tilde{\mathbf{Z}}).
\label{eq:priv_loss}
\end{equation*}
We optimize only server-side components, training the decoder to interpret obfuscated embeddings for domain tasks.

\subsection{Inference: Text-free Prompting at Runtime}
\label{sec:inference}

At inference time, the client encodes the prompt, applies $k$-pooling and noise injection, and transmits only $\tilde{\mathbf{U}}$.
The server projects $\tilde{\mathbf{U}}$ to $\tilde{\mathbf{Z}}$ and generates $\mathbf{y}$ with the fine-tuned decoder, so the prompt text never leaves the device.

\begin{table*}[t]
\centering
\small
\renewcommand{\arraystretch}{1.15}
\begin{tabular}{ll|c|ccc}
\toprule
\textbf{Backbone} &
\textbf{Method} &
\textbf{Average} &
\textbf{Pri-DDX} & \textbf{Pri-NLICE} & \textbf{Pri-SLJA} \\
\midrule
\multirow{6}{*}{\shortstack{Llama-3.1-8B}}
  & $d_\chi$-privacy \cite{feyisetan2020privacy}        & 0.2750 \dec{0.6541} & 0.2311 & 0.3477 & 0.2462 \\
  & Paraphrase \cite{utpala2023locally}              & 0.3757 \dec{0.5534} & 0.4648 & 0.2892 & 0.3731 \\
  & PrivacyRestore \cite{zeng2025privacyrestore} & 0.6343 \dec{0.2948} & 0.5784 & 0.5415 & 0.7829 \\
  & \cellcolor{gray!20}\textbf{PPFT (Ours)}
    & \cellcolor{gray!20}\textbf{0.7314} \dec{0.1977}
    & \cellcolor{gray!20}\textbf{0.5915}
    & \cellcolor{gray!20}\textbf{0.6979}
    & \cellcolor{gray!20}\textbf{0.9049} \\
  & \quad \textit{$\text{PPFT}_{\text{w/o stage2}}$ (Lower Bound)} & \textit{0.3545} & \textit{0.3460} & \textit{0.3138} & \textit{0.4036} \\
  & \quad \textit{$\text{PPFT}_{\text{w/o noise}}$ (Upper Bound)}    & \textit{0.9291} & \textit{0.9275} & \textit{0.9049} & \textit{0.9466}  \\
\midrule

\multirow{6}{*}{\shortstack{Llama-3.2-1B}}
  & $d_\chi$-privacy \cite{feyisetan2020privacy}       & 0.2608 \dec{0.4965} & 0.3176 & 0.2631 & 0.2018 \\
  & Paraphrase \cite{utpala2023locally}             & 0.2635 \dec{0.4938} & 0.2382 & 0.1753 & 0.3770 \\
  & PrivacyRestore  \cite{zeng2025privacyrestore}& 0.4519 \dec{0.3054} & \textbf{0.5150} & 0.4277 & 0.4128 \\
  & \cellcolor{gray!20}\textbf{PPFT (Ours)}
    & \cellcolor{gray!20}\textbf{0.5699} \dec{0.1874}
    & \cellcolor{gray!20}0.4537
    & \cellcolor{gray!20}\textbf{0.4866}
    & \cellcolor{gray!20}\textbf{0.7693} \\
  & \quad \textit{$\text{PPFT}_{\text{w/o stage2}}$ (Lower Bound)} & \textit{0.3788} & \textit{0.3707} & \textit{0.3008} & \textit{0.4648}  \\
  & \quad \textit{$\text{PPFT}_{\text{w/o noise}}$ (Upper Bound)}    & \textit{0.7573} & \textit{0.7071} & \textit{0.6622} & \textit{0.9003} \\
\bottomrule
\end{tabular}
\caption{Main results on downstream tasks. \textbf{PPFT ($k=4$)} refers to our model adapted with noise in Stage 2. Lower/Upper bounds indicate performance without domain adaptation and without privacy noise, respectively.}
\label{tab:main_results}
\end{table*}

\section{Experiments}\label{sec:experiments}

\subsection{Experimental Setup}
\label{sec:exp_setup}

We evaluate PPFT under text-free operation along two axes: (i) downstream task performance and (ii) robustness to prompt reconstruction (inversion) attacks.

\paragraph{Models and Training Stages.}
We adopt ModernBERT-large~\cite{warner2025smarter} as the client-side encoder, chosen for its strong embedding quality while remaining lightweight enough to run efficiently on commodity client hardware (CPU-only) without requiring a dedicated accelerator.
For the server-side decoder, we use Llama-3.2-1B-Instruct and Llama-3.1-8B-Instruct to examine scaling behavior across model sizes~\cite{dubey2024llama}.
All hyperparameters are provided in Appendix~\ref{app:training_details}.

\paragraph{Datasets.}
Stage~1 uses general-domain data for interface alignment, while Stage~2 uses medical and legal QA datasets to reflect sensitive-domain adaptation~\cite{zeng2025privacyrestore}. Data sources and preprocessing are described in Appendix~\ref{app:datasets}.

\paragraph{Baselines and Reference Points.}
We compare against major prompt-protection paradigms: representation perturbation ($d_\chi$-privacy)~\cite{feyisetan2020privacy}, text transformation (Paraphrase)~\cite{utpala2023locally}, and reconstruction-evaluation frameworks (PrivacyRestore)~\cite{zeng2025privacyrestore}.
We also report two reference points. \emph{Stage~1 only} serves as a \emph{lower bound} because it uses the text-free interface \emph{without} domain adaptation. \emph{Stage~2 without noise} serves as an \emph{upper bound} because it follows the same pipeline and supervision but removes privacy noise, approximating the best achievable performance under our interface.
Implementation details and ablations are deferred to Appendix~\ref{app:Baseline Details}.

\paragraph{Evaluation}
We separately evaluate (i) domain performance via downstream task accuracy and (ii) privacy robustness via reconstruction resistance.
For downstream tasks, a prediction is counted as correct if the generated output contains the normalized gold answer text, following standard MCQA and extractive QA evaluation practice.
Privacy robustness is assessed by measuring how well an attacker can reconstruct the original prompt from transmitted embeddings using ROUGE-L, where lower scores indicate stronger resistance.
Task-specific metrics, scoring rules, and privacy evaluation procedures are detailed in Appendix~\ref{app:Evaluation_Metrics}.

\paragraph{Privacy Budget Analysis and Fair Comparison}
For fair comparison, we align privacy budgets across all methods under a unified $d_\chi$-privacy accounting; the resulting calibration and $\epsilon$ settings are reported in Appendix~\ref{app:privacy_budget} and Appendix~\ref{app:privacy_hparams}.


\begin{table*}[t]
\centering
\small
\setlength{\tabcolsep}{6pt}
\renewcommand{\arraystretch}{1.25}
\begin{tabular}{p{0.96\textwidth}}
\hline
\textbf{Original Prompt} \\
A 27-year-old male has a history of chronic pancreatitis, diabetes, obesity, pancreatic cancer in family members, smoking.\\
The 27-year-old male presents the symptoms of diarrhea, fatigue, nausea, pain, pale stools and dark urine, skin lesions, underweight.\\
What is the likely diagnosis? \\
\hline
\textbf{Reconstructed by Inversion Attack (same $\epsilon$ as inference)} \\
\same{A }\diff{28}\same{-year-old }\diff{woman }\same{has a history of }
\diff{asthma, asthma attack, asthma attack, asthma attack, asthma attack, asthma attack, asthma attack, asthma attack, asthma attack, asthma attack, asthma attack, asthma attack.}\\
\same{The }\diff{28}\same{-year-old }\diff{woman }\same{presents the symptoms of }
\diff{cough, wheezing, shortness of breath, shortness of breath, wheezing, shortness of breath, shortness of breath with deep breathing.}\\
\same{What is the likely diagnosis?} \\
\hline
\end{tabular}
\caption{Qualitative reconstruction example under noisy-embedding transmission.
\textcolor{blue}{Blue} indicates spans that exactly match the original prompt, whereas \textcolor{red}{red} indicates mismatched content.}
\label{tab:recon_example}
\end{table*}

\subsection{Main Results: Domain Performance}
\label{sec:main_results}

We evaluate whether PPFT preserves domain performance under strict text-free constraints on medical and legal test sets.
We compare PPFT against the lower bound, the noise-free upper bound, and competing privacy-preserving baselines under identical evaluation conditions. As shown in Table~\ref{tab:main_results}, PPFT achieves the best overall task performance with the 8B decoder across all datasets and baselines. 
With the 1B decoder, PPFT remains top-performing on all benchmarks except Pri-DDX, indicating that strong performance can be preserved even under a fully text-free training and inference interface.
Notably, on the legal-domain Pri-SLJA dataset, PPFT with noise injection recovers performance close to the noise-free upper bound ($\text{PPFT}_{\text{w/o noise}}$), achieving 95.6\% task accuracy with the 8B model and 85.0\% with the 1B model. This indicates that PPFT preserves most domain-critical semantics despite operating under strong privacy constraints.

We can also observe that baseline methods exhibit distinct failure modes.
$d_\chi$-privacy frequently distorts symptom expressions or sentence structure through word-level noise and nearest-neighbor substitutions, altering clinical semantics and hindering correct answer selection.
Paraphrasing often replaces or omits key diagnostic cues during rewriting, leading to reduced accuracy.
PrivacyRestore struggles to recover domain-critical semantics from masked representations, resulting in downstream performance loss. In contrast, PPFT performs privacy protection entirely at the embedding level without modifying text.
Since the decoder directly adapts to obfuscated embeddings during Stage~2, PPFT consistently retains domain performance close to the upper bound. Overall, PPFT limits the degradation from the upper bound to below 0.2 while maintaining competitive domain adaptation without ever exposing prompt text to the server. These results clearly demonstrate the effectiveness of PPFT.

\begin{figure}[t]
    \centering
    \includegraphics[width=1.0\linewidth]{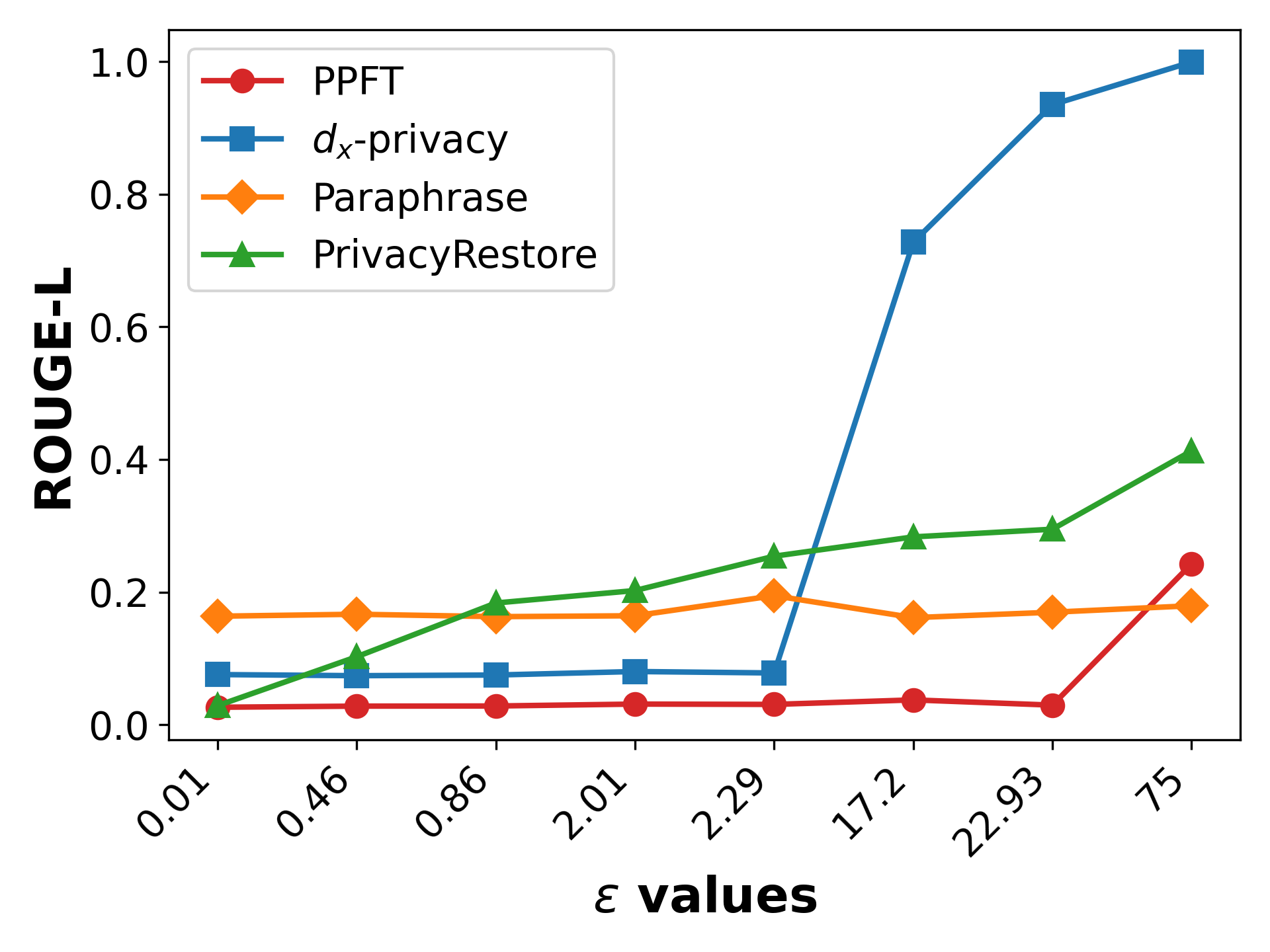}
    \caption{Results of embedding inversion attacks and attribute inference attacks across all baselines under varying privacy budgets $\epsilon$ on Pri-DDX.}
    \label{fig:noise_scale}
\end{figure}

\subsection{Reconstruction Resistance under Inversion Attacks}
\label{sec:privacy_results}

We assess PPFT robustness against inversion attacks that attempt to reconstruct original prompts from observable embeddings, reflecting a realistic threat model in embedding-based transmission settings.
The attacker first pretrains a reconstruction model using clean embeddings and then evaluates reconstruction quality on obfuscated embeddings using ROUGE-L as the similarity metric.
Attack architectures, training protocols, and evaluation details are provided in Appendix~\ref{app:Inverse Attack}.

Figure~\ref{fig:noise_scale} reports reconstruction performance across noise scale $\epsilon$.
As expected, reconstruction accuracy generally increases with larger $\epsilon$ (weaker noise).
However, PPFT consistently maintains low ROUGE-L scores across a wide range of $\epsilon$ values, indicating strong resistance even under powerful adversarial settings. While paraphrasing may appear favorable under reconstruction metrics because it directly alters text, this comes at the cost of semantic distortion. 
PPFT, in contrast, preserves textual semantics by operating entirely under text-free constraints and injecting noise only at the continuous embedding level.
Even at $\epsilon{=}75$, PPFT keeps ROUGE-L below 0.25, achieving a practical level of privacy protection.

This trend remains consistent under the stronger attacker settings in Appendix~\ref{app:inverse_noise_aware}, Appendix~\ref{app:inverse_stage1_aligned}, and Appendix~\ref{app:pooling_inversion}.

\paragraph{Qualitative analysis of reconstruction.}
Table~\ref{tab:recon_example} presents qualitative examples of inversion attack outputs from obfuscated embeddings.
While reconstructed text may partially preserve surface structure, core semantic slots collapse into repetitive or incoherent content.
These observations qualitatively support that PPFT’s noise injection substantially impedes recovery of sensitive clinical information, even when superficial text patterns remain.

\begin{table}[t]
\centering
\small
\setlength{\tabcolsep}{6pt}
\resizebox{\linewidth}{!}{%
\begin{tabular}{lcccc}
\toprule
\textbf{Method} 
& \textbf{Age} 
& \textbf{Sex} 
& \textbf{Symptom} 
& \textbf{Antecedent} \\
\midrule
PrivacyRestore              & - & \textbf{0.5642} & 0.3552 & 0.3317 \\
\textbf{PPFT (Ours)}        & \textbf{0.0071} & 0.5894 & \textbf{0.1001} & \textbf{0.0115} \\
\bottomrule
\end{tabular}}
\caption{Fine-grained reconstruction error on the Pri-DDX dataset under inference-level privacy budgets.}
\label{tab:inv_finegrained_recon}
\end{table}

\paragraph{Attribute-level analysis of inversion attacks.}
We analyze inversion attacks using attribute-level \emph{recall} over four sensitive attributes—age, sex, current symptoms, and prior antecedents—where lower recall indicates weaker recovery of private information.
All experiments are conducted on the Pri-DDX dataset under the same privacy budget $\epsilon$ as used during inference.
As shown in Table~\ref{tab:inv_finegrained_recon}, PPFT exhibits consistently low recall across all attributes, indicating that sensitive information is largely not reconstructed.
In particular, age(0.0071) and Antecedent(0.0115) are almost never recovered, while sex recall (0.5894) remains close to a random baseline for a binary attribute (0.5).

In contrast, PrivacyRestore achieves higher recall than PPFT on all attributes except sex.
While PrivacyRestore masks symptoms and antecedents and provides age and sex as inputs, it yields only about 57\% exact-match correctness on these demographic fields, yet still exhibits substantially higher reconstruction recall for current symptoms (0.3552) and prior antecedents (0.3317).
This indicates that despite preserving demographic consistency, PrivacyRestore fails to prevent the recovery of medically sensitive content.
Overall, these results show that high ROUGE-L scores primarily reflect imitation of surface-level clinical templates, whereas PPFT effectively prevents the reconstruction of underlying private attributes that define the sensitive medical context.

\begin{table}[t]
\centering
\small
\resizebox{\columnwidth}{!}{%
\renewcommand{\arraystretch}{1.15}
\begin{tabular}{ll|cc}
\toprule
\textbf{Backbone} & \textbf{Method} & \textbf{CSQA} & \textbf{SQuAD} \\
\midrule
\multirow{4}{*}{\shortstack{Llama-3.1-8B}} 
 & $d_\chi$-privacy & 0.1819 & 0.0174 \\
 & Paraphrase & 0.0649 & 0.0125 \\
 & \cellcolor{gray!20}\textbf{PPFT (Ours)} & \cellcolor{gray!20}\textbf{0.5278} & \cellcolor{gray!20}\textbf{0.7085} \\
  & \quad \textit{$\text{PPFT}_{\text{w/o noise}}$} & \textit{0.6086} & \textit{0.8930} \\
\midrule
\multirow{4}{*}{\shortstack{Llama-3.2-1B}} 
 & $d_\chi$-privacy & 0.1210 & 0.0313 \\
 & Paraphrase & 0.0470 & 0.072 \\
 & \cellcolor{gray!20}\textbf{PPFT (Ours)} & \cellcolor{gray!20}\textbf{0.5125} & \cellcolor{gray!20}\textbf{0.6579} \\
  & \quad \textit{$\text{PPFT}_{\text{w/o noise}}$} & \textit{0.543} & \textit{0.7303} \\
\bottomrule
\end{tabular}%
}
\caption{Performance on general domains.}
\label{tab:general_result}
\end{table}

\subsection{General-domain Performance}
\label{sec:general_results}

We evaluate whether injecting noise during privacy-preserving fine-tuning degrades general-domain performance. 
To isolate the effect of noise, we use $\text{PPFT}_{\text{w/o noise}}$ as the reference baseline and measure the performance drop incurred when noise is introduced under an otherwise identical training and inference interface.  

Table~\ref{tab:general_result} reports results on general-domain benchmarks.
Across model scales, PPFT exhibits only limited degradation relative to the noise-free baseline.
For the LLaMA-3.1-8B model, performance drops are modest, with decreases of 0.081 on CSQA and 0.184 on SQuAD.
Notably, the LLaMA-3.2-1B model shows even smaller losses, incurring reductions of only 0.030 on CSQA and 0.072 on SQuAD.

In contrast, $d_\chi$-privacy and Paraphrase frequently corrupt information critical for answer selection, leading to significant systematic errors.
Despite being adapted exclusively on sensitive-domain data without additional general-domain replay, PPFT maintains robust general reasoning.
This robustness can be attributed to the two-stage design: Stage~1 establishes a stable text-free alignment between embeddings and the decoder, while Stage~2 introduces noise-aware adaptation without disrupting the model’s general capabilities.

\section{Ablation Study}
This section examines how key design choices in PPFT shape the trade-off between task performance and privacy protection.
Specifically, we analyze (i) the effect of the pooling size $k$ on downstream performance and reconstruction resistance, highlighting the performance--privacy trade-off induced by different levels of token compression, and (ii) the impact of noise design, comparing different noise mechanisms as well as the no-noise setting to quantify their relative effectiveness in mitigating reconstruction attacks.


\begin{table}[h]
\centering
\small
\resizebox{0.8\columnwidth}{!}{%
\renewcommand{\arraystretch}{1.15}
\begin{tabular}{l|ccc}
\toprule
\multirow{2}{*}{\textbf{Metric}} & \multicolumn{3}{c}{\textbf{Pooling Size $(k)$}} \\
 & \textbf{4} & \textbf{8} & \textbf{16} \\
\midrule
\textbf{Score$\uparrow$} & \textbf{0.9049} & 0.8363 & 0.7630 \\
\textbf{ROUGE-L$\downarrow$} & 0.4050 & 0.3553 & \textbf{0.3241} \\
\bottomrule
\end{tabular}%
}
\caption{Ablation study on pooling size $k$. ROUGE-L is measured on the Pri-SLJA test set.}
\label{tab:pooling_result}
\end{table}

\subsection{Effect of Pooling Size $k$}
\label{sec:pooling_results}

Table~\ref{tab:pooling_result} reports the trade-off between domain performance and reconstruction ease (measured by ROUGE-L) as the pooling size $k$ varies.
All ROUGE-L scores are computed under the same privacy setting ($\epsilon{=}75$) using an inversion-based reconstruction model, and we evaluate this ablation on the Pri-SLJA test set.
When $k{=}4$, PPFT preserves the highest domain performance; however, ROUGE-L is also relatively high, indicating that embeddings retain more recoverable information.
As $k$ increases, the input representation is more aggressively compressed, leading to a gradual decline in task performance, while ROUGE-L consistently decreases, indicating \textbf{stronger resistance to reconstruction attacks}.
We note that ROUGE-L values on Pri-SLJA can appear relatively high in absolute terms because many samples share a long, standardized legal instruction prefix, making partial-prefix recovery easier even when the remainder of the prompt is poorly reconstructed.

Overall, the pooling size $k$ acts as a key control knob that jointly regulates communication efficiency and the performance--privacy balance.


\subsection{Effect of Noise Types}
\label{sec:noise_ablation}

Figure~\ref{fig:noise_type} compares reconstruction resistance across noise types.
With Gaussian noise, ROUGE-L exceeds 0.2 even at low privacy budgets $\epsilon$, suggesting that embeddings remain relatively vulnerable to generative inversion attacks. In contrast, Laplace noise consistently yields lower ROUGE-L across all $\epsilon$ values. Although reconstruction performance gradually increases as $\epsilon$ grows, Laplace noise provides stronger overall resistance than its Gaussian counterpart.

This behavior suggests that Laplace noise more effectively degrades semantic reconstructability in high-dimensional embedding spaces.

\subsection{Effect of Noise Injection}
\label{sec:noise_on_off}

Beyond noise type, we examine whether reconstruction resistance primarily arises from noise injection itself. We directly compare settings with no noise and with noise injected at the same $\epsilon$ used during inference under otherwise identical conditions.

\begin{figure}[t]
  \centering
  \begin{minipage}[t]{0.48\linewidth}
    \centering
    \includegraphics[width=\linewidth]{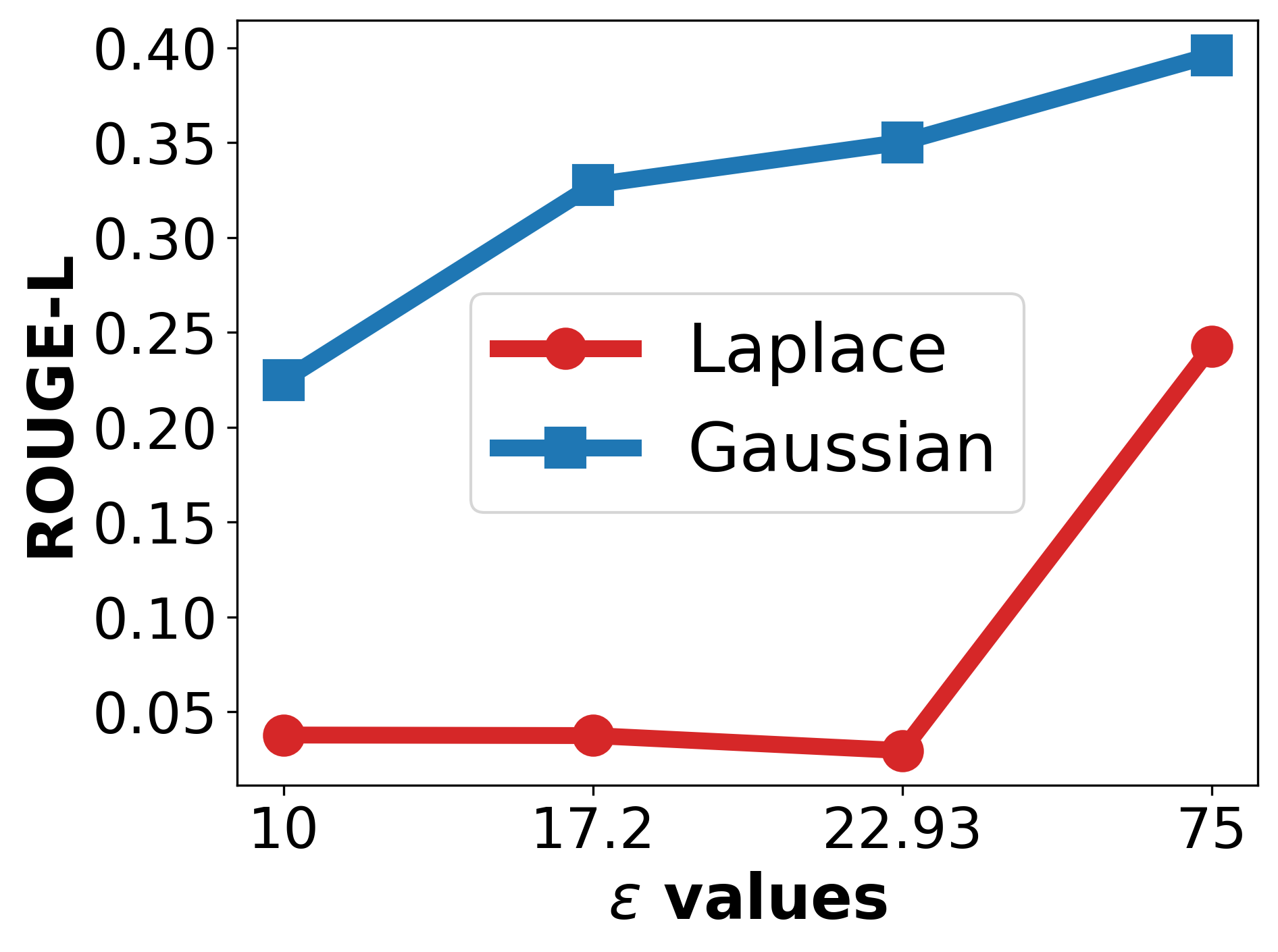}
    \captionof{figure}{Reconstruction performance under different noise types.}
    \label{fig:noise_type}
  \end{minipage}\hfill
  \begin{minipage}[t]{0.48\linewidth}
    \centering
    \includegraphics[width=\linewidth]{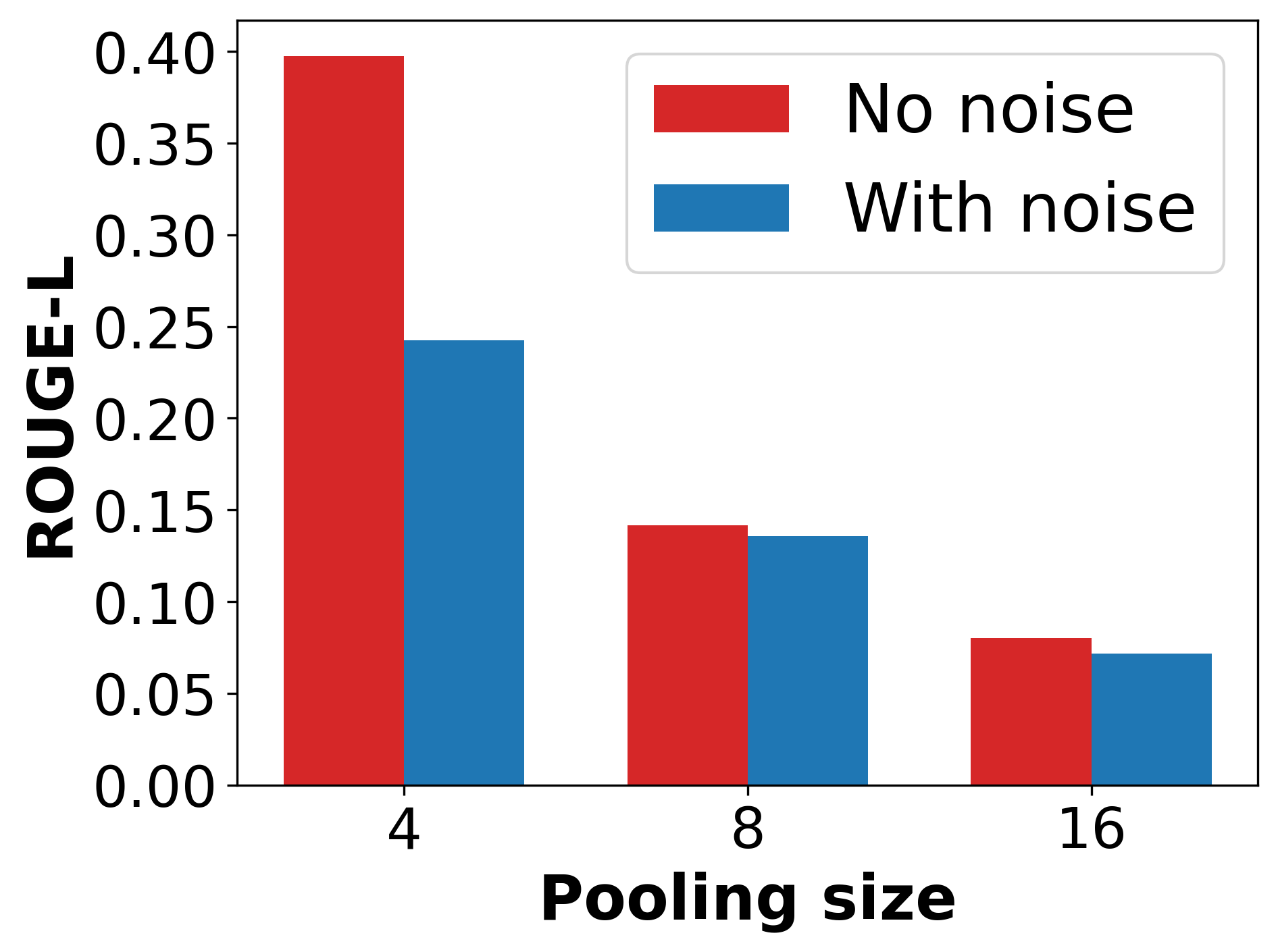}
    \captionof{figure}{Reconstruction performance with and without noise injection.}
    \label{fig:noise_no_noise}
  \end{minipage}
\end{figure}

As shown in Figure~\ref{fig:noise_no_noise}, noise injection consistently reduces ROUGE-L across all pooling sizes $k$, thereby increasing reconstruction difficulty and strengthening privacy protection. The effect is most pronounced at $k{=}4$, where embeddings retain higher information content. This observation indicates that noise injection plays a particularly critical defensive role when embeddings are less compressed.

\section{Conclusion}
In this paper, we propose PPFT (Privacy-Preserving Fine-Tuning), a framework that ensures \textbf{prompt text never becomes visible to the server during either inference or domain-specific fine-tuning} in the post–pre-training stage of LLMs. PPFT fundamentally blocks text transmission by converting prompts into continuous embeddings on the client side. It further applies $k$-Pooling to aggregate token representations, intentionally lowering the information resolution of input sequences to impede the reconstruction of fine-grained token details. We additionally integrate $d_\chi$-privacy–based noise injection, which effectively suppresses generative inversion attacks that attempt to recover original prompts from observable embeddings.

Empirically, PPFT consistently outperforms existing privacy-preserving baselines—including $d_\chi$-privacy, paraphrasing, and PrivacyRestore—across medical and legal domains. While incurring only limited performance degradation relative to a noise-free upper bound, PPFT achieves substantially lower reconstruction scores (ROUGE-L) under strong inversion attacks. Notably, even under strict text-free constraints, PPFT recovers up to approximately 95\% of the upper-bound utility, demonstrating its practicality for real-world deployment. These results indicate that PPFT provides a scalable and effective solution for MLaaS environments where privacy and performance must be balanced without exposing raw data.

\section*{Limitations}
We identify potential privacy risks in LLM-based services and propose an effective mitigation strategy. Within the scope of our proposal, we conducted rigorous validation and provided sufficient empirical evidence to support our claims. However, due to resource and page-limit constraints, we do not address all possible privacy issues. We summarize the limitations of our study as follows.

\paragraph{Output-side exposure.}
PPFT strengthens \emph{input confidentiality} by ensuring that prompt text never reaches the server during inference or fine-tuning. However, because model outputs must ultimately be delivered to users, PPFT does not structurally prevent the exposure of generated content itself. As a result, PPFT guarantees \emph{prompt non-disclosure} rather than end-to-end content confidentiality. In practical deployments, PPFT should therefore be complemented with output-side safeguards such as content filtering, policy-based controls, and sensitive information detection or masking mechanisms.

\paragraph{Generality across model pairs and modalities.}
We validate PPFT using a ModernBERT-large encoder paired with LLaMA-family decoders in text-based medical and legal domains. Whether the same continuous-embedding input interface can be efficiently supported by smaller client-side encoders, alternative decoder architectures, or closed-source API-based LLMs requires further investigation. In addition, extending PPFT to multilingual or multimodal inputs raises open questions about whether the same utility–privacy trade-offs can be preserved across modalities.

\section*{Ethics Statement}

\paragraph{Data sources and licensing.}
All experiments in this paper use \emph{publicly available} datasets. We do not collect any new data involving human subjects, nor do we attempt to identify any individual.

\paragraph{Personally identifying information (PII) and offensive content checks.}
The primary sensitive-domain datasets used in our study (the \textsc{Pri} datasets) are taken from prior work~\cite{zeng2025privacyrestore}. These datasets are \emph{synthetically generated} and are designed to contain \emph{fictional individuals} rather than real persons. As a result, the datasets are not expected to include real-world personally identifying information. In addition, we treat the \textsc{Pri} datasets as sensitive by design (e.g., clinical/legal style content) and adopt conservative handling: we do not release any raw prompts beyond what is already publicly available, and we avoid exposing original prompt text in our proposed text-free interface.

\paragraph{Data protection and anonymization.}
Although the \textsc{Pri} datasets are synthetic, we follow the spirit of privacy-preserving research by minimizing exposure of potentially sensitive attributes. In PPFT, the client never transmits prompt text to the server; instead, the server only receives compressed and noise-injected continuous representations. This design further reduces the risk of leaking user-provided content during both inference and fine-tuning.



\bibliography{section/99references}

\appendix


\section{Training Details}
\label{app:training_details}

Our architecture consists of an encoder and a decoder.
For the encoder, we use \href{https://huggingface.co/answerdotai/ModernBERT-large}{\texttt{answerdotai/ModernBERT-large}}~\cite{warner2025smarter}, while the decoder is instantiated from instruction-tuned LLaMA models~\cite{dubey2024llama}.
Specifically, we evaluate two decoder backbones:
\href{https://huggingface.co/meta-llama/Llama-3.2-1B-Instruct}{\texttt{meta-llama/Llama-3.2-1B-Instruct}} and
\href{https://huggingface.co/meta-llama/Llama-3.1-8B-Instruct}{\texttt{meta-llama/Llama-3.1-8B-Instruct}}.
Unless otherwise stated, we apply the same training configuration across model scales to ensure fair comparison.

\paragraph{Model Configuration.}
The maximum sequence length is set to 512 tokens for both the encoder and decoder.
We apply Low-Rank Adaptation (LoRA) to the decoder, with rank $r=16$ and scaling factor $\alpha=32$.

\paragraph{Optimization.}
We use the AdamW optimizer with a cosine learning rate schedule and a warmup ratio of 0.1.
The peak learning rate is set to $2\times10^{-5}$ for both Stage~1 and Stage~2.

\paragraph{Stage-specific Settings.}
Stage~1 (alignment) and Stage~2 (domain adaptation) share identical optimization hyperparameters.
In Stage~2, we reduce the per-device batch size from 8 to 4 in order to increase the number of optimization steps per epoch, allowing the model to better adapt to the injected noise during privacy-preserving training.
A complete summary of hyperparameters is provided in Table~\ref{tab:hyperparameters}.

\begin{table}[h]
\centering
\small
\setlength{\tabcolsep}{8pt}
\renewcommand{\arraystretch}{1.0}
\resizebox{\linewidth}{!}{%
\begin{tabular}{lc}
\toprule
\textbf{Hyperparameter} & \textbf{Value} \\
\midrule
\multicolumn{2}{c}{\textit{General Settings}} \\
Backbones & Llama-3.2-1B / 3.1-8B \\
Precision & bfloat16 \\
Max Sequence Length & 512 \\
\midrule
\multicolumn{2}{c}{\textit{LoRA Configuration}} \\
Rank ($r$) & 16 \\
Alpha ($\alpha$) & 32 \\
Dropout & 0.05 \\
\midrule
\multicolumn{2}{c}{\textit{Optimization (AdamW)}} \\
Peak Learning Rate & 2e-5 \\
Weight Decay & 0.01 \\
Beta1, Beta2 & 0.9, 0.999 \\
Epsilon & 1e-8 \\
Scheduler & Cosine \\
Warmup Ratio & 0.1 \\
\midrule
\multicolumn{2}{c}{\textit{Stage 1 Specifics}} \\
Epochs & 1 \\
Batch Size & 8 \\
Gradient Accumulation & 1 \\
\midrule
\multicolumn{2}{c}{\textit{Stage 2 Specifics}} \\
Batch Size & 4 \\
(Other params same as Stage 1) & -- \\
\bottomrule
\end{tabular}}
\caption{Hyperparameters used for training Llama-3.2-1B and Llama-3.1-8B models across Stage 1 and Stage 2.}
\label{tab:hyperparameters}
\end{table}

\section{Dataset Details}
\label{app:datasets}

\subsection{Overview.}
We use a two-stage training pipeline: Stage~1 (general-domain alignment) and Stage~2 (domain adaptation under the text-free interface).
All datasets are converted into a unified instruction-following format with consistent field ordering and a shared length constraint.

\subsection{Stage~1: General-Domain Alignment Corpora}
Stage~1 trains the model to generate answers from continuous prefix embeddings using general-domain instruction and QA data.
\begin{itemize}[leftmargin=1.2em, itemsep=0.1em]
  \item \href{https://huggingface.co/datasets/allenai/ai2_arc}{\texttt{allenai/ai2\_arc}}~\cite{clark2018think}
  \item \href{https://huggingface.co/datasets/TIGER-Lab/WebInstructSub}{\texttt{TIGER-Lab/WebInstructSub}}~\cite{yue2024mammoth2}
  \item \href{https://huggingface.co/datasets/yahma/alpaca-cleaned}{\texttt{yahma/alpaca-cleaned}}~\cite{taori2023stanford}
  \item \href{https://huggingface.co/datasets/databricks/databricks-dolly-15k}{\texttt{databricks/databricks-dolly-15k}}~\cite{conover2023free}
  \item \href{https://huggingface.co/datasets/nvidia/ChatQA-Training-Data}{\texttt{nvidia/ChatQA-Training-Data}}~\cite{liu2024chatqa} (SFT split)
  \item \href{https://huggingface.co/datasets/rajpurkar/squad}{\texttt{rajpurkar/squad}}~\cite{rajpurkar2016squad}
  \item \href{https://huggingface.co/datasets/tau/commonsense_qa}{\texttt{tau/commonsense\_qa}}~\cite{talmor2019commonsenseqa}
\end{itemize}

\subsection{Stage~2: Domain Adaptation Corpora}
Stage~2 adapts the aligned model to sensitive domains (medical and legal) while preserving the text-free training interface.
To strengthen MCQA behavior for both decoders, we additionally include \href{https://huggingface.co/datasets/pszemraj/unified-mcqa}{\texttt{pszemraj/unified-mcqa}}.

\paragraph{Medical.}
{\small
\begin{itemize}[leftmargin=1.2em, itemsep=0.1em]
  \item \href{https://huggingface.co/datasets/openlifescienceai/medmcqa}{\texttt{openlifescienceai/medmcqa}}~\cite{pal2022medmcqa}
  \item \href{https://huggingface.co/datasets/medalpaca/medical_meadow_medical_flashcards}{\texttt{medalpaca/medical\_meadow\_medical\_flashcards}}
  \item \texttt{Pri-NLICE}, \texttt{Pri-DDX} (constructed following \textsc{PrivacyRestore};
  \href{https://github.com/wjw136/PrivacyRestore}{GitHub})
\end{itemize}
}
\paragraph{Legal.}
{\small
\begin{itemize}[leftmargin=1.2em, itemsep=0.1em]
  \item \href{https://huggingface.co/datasets/ramo6627/open-australian-legal-qa-formatted-2k}{\texttt{ramo6627/open-australian-legal-qa-formatted-2k}}
  \item \href{https://huggingface.co/datasets/dzunggg/legal-qa-v1}{\texttt{dzunggg/legal-qa-v1}}
  \item \texttt{Pri-SLJA} (constructed under the same pipeline)
\end{itemize}
}

\subsection{Unified prompt construction.}
Each example is serialized into a single input string by concatenating available fields in a fixed order:
\emph{instruction}, \emph{context}, and \emph{question}.
If an instruction is present, we prepend it as ``\texttt{instruction: ...}''.
If a context is present, we append it as ``\texttt{context: ...}''.
For the question, we use ``\texttt{question: ...}'' only when an instruction and/or context exists; otherwise, we use the raw question text.
The final training target is the corresponding answer string.

\subsection{Length filtering.}
We discard examples whose concatenated \texttt{(input + answer)} exceeds 512 tokens under the decoder tokenizer,
to keep training stable and to match practical deployment constraints.

\subsection{MCQA normalization.}
\label{app:MCQA}
For all MCQA-style datasets (including training and test sets), we prepend a standardized instruction:
\begin{quote}
\small \texttt{Choose the correct option and output only its text, not the label.}
\end{quote}
Options are appended using an ``\texttt{options: ...}'' block.
This normalization is critical in our setting because compression (via pooling) can preserve semantic content
while weakening the correspondence between option labels (e.g., A/B/C/D) and option texts.
Accordingly, we evaluate and train models to output the \emph{option text} rather than the label.

\section{Baseline Details}
\label{app:Baseline Details}

This appendix describes the baselines and reference configurations used throughout our experiments.
Unless otherwise noted, all baselines are evaluated under the same MCQA inference protocol described in
Appendix~\ref{app:MCQA}.
For a fair comparison, \textbf{only the question (and its associated context, if any) is obfuscated};
the \emph{MCQA instruction} and \emph{options block} are kept unchanged (i.e., not perturbed) for all methods.

\subsection{PPFT Upper/Lower Bounds}

\paragraph{PPFT without noise (Upper Bound).}
This configuration starts from the Stage~1 aligned PPFT model and performs Stage~2 domain adaptation
\emph{without} applying any privacy noise to the client-side embeddings.
Since the training interface and optimization remain identical while removing the privacy constraint,
this setting provides an approximate \emph{upper bound} on task performance.
Empirically, it achieves the best domain performance and preserves general-domain capabilities more strongly than privacy-constrained variants.

\paragraph{PPFT without Stage~2 (Lower Bound).}
This configuration evaluates the Stage~1 aligned model directly on the domain-specific test sets
\emph{without} any Stage~2 domain adaptation.
Because Stage~1 uses only general-domain corpora, the model lacks domain knowledge required for medical/legal QA,
leading to substantially worse in-domain performance while retaining relatively strong general-domain behavior.
We report this setting as a \emph{lower bound} for domain adaptation.

\subsection{Token-level Perturbation Baseline: $d_\chi$-privacy}
\paragraph{$d_\chi$-privacy (word-level privatization).}
Following \citet{feyisetan2020privacy}, we apply a token-level privatization mechanism based on $d_\chi$-privacy.
Specifically, each token in the user query is independently replaced by a randomized alternative sampled from the vocabulary according to a distance-based distribution defined in a semantic embedding space.
The sampling probability decays exponentially with the distance from the original token, ensuring $d_\chi$-privacy at the word level.
The resulting obfuscated text query is then sent to the server for inference or fine-tuning, depending on the setting.
For the underlying semantic space used to compute token distances, we employ \texttt{glove.840B.300d} embeddings.

\subsection{Generative Text Privatization Baseline: Paraphrase}
\paragraph{Paraphrase.}
\citet{utpala2023locally} argue that token-level privatization methods may incur privacy-budget growth as input length increases, and propose paraphrasing via a generative model as a text-based privacy baseline.
Such approaches aim to obfuscate sensitive content by rephrasing the input while preserving task-relevant semantics, without providing formal differential privacy guarantees.
In our experiments, to reflect realistic client-side compute constraints and to use a model of comparable scale to our client encoder, we employ \href{https://huggingface.co/google/flan-t5-base}{\texttt{google/flan-t5-base}}~\cite{chung2024scaling} on the client side to generate paraphrases.
We prompt the paraphraser with:
\begin{quote}
\small \texttt{Paraphrase this sentence while hiding personal information.}
\end{quote}
The paraphrased query is then used for downstream inference or training under the same protocol as other baselines.

\subsection{Recovery-based Baseline: PrivacyRestore}
\paragraph{PrivacyRestore.}
We compare against PrivacyRestore~\cite{zeng2025privacyrestore}, which studies the trade-off between privacy protection and utility under masked personally identifiable information (PII).
PrivacyRestore introduces a recovery mechanism based on auxiliary representations (e.g., meta vectors) to partially reconstruct masked content when needed.
In our evaluation, we follow the original PrivacyRestore setup to generate masked inputs and apply its recovery procedure, and then perform downstream inference using the recovered (or partially recovered) queries under the same MCQA pipeline as other methods (Appendix~\ref{app:MCQA}).

\paragraph{Inference protocol (shared).}
All baselines and PPFT variants are evaluated under the same MCQA formatting and decoding rules (Appendix~\ref{app:MCQA}).
Privacy transformations are applied only to the question (and context), while the instruction and answer options remain unchanged to ensure a fixed decision interface across methods.

\section{Evaluation Metrics}
\label{app:Evaluation_Metrics}

We report two complementary metrics: (i) task performance measured by accuracy on downstream QA tasks, and
(ii) privacy / reconstruction resistance measured by ROUGE-L under inversion attacks.
All reported results are obtained from a single evaluation run per configuration.

\subsection{Downstream Utility: Accuracy}
\label{app:acc_metric}

We measure downstream task performance using accuracy.
Under the MCQA setup (Appendix~\ref{app:MCQA}), a prediction is considered correct if the model outputs the gold option text after normalization.
We evaluate option texts rather than option labels to ensure consistency across different privatization and compression settings.

\subsection{Reconstruction Resistance: ROUGE-L}
\label{app:ROUGE_metric}

For inversion attacks, we evaluate how well an attacker can reconstruct the original user prompt from transmitted embeddings.
We measure reconstruction quality using ROUGE-L~\cite{lin2004ROUGE}, which is based on the length of the
Longest Common Subsequence (LCS) between the reconstructed text and the original text.
ROUGE-L captures both token overlap and sequence-level ordering, making it suitable for detecting whether an attacker
recovers substantial portions of the original prompt (including key entities and symptom descriptions) in the correct structure.
Lower ROUGE-L indicates stronger reconstruction resistance (i.e., better privacy protection).


\section{Privacy Budget and Alignment Rules}
\label{app:privacy_budget}
A critical challenge in comparing privacy-preserving mechanisms for LLMs is ensuring a fair alignment between methods that operate on different granularities (e.g., tokens vs.\ embeddings) and composition rules. To address this, we align all baselines and our method (PPFT) to a unified Global Privacy Budget ($B$), rather than comparing local $\epsilon$ values in isolation.

\subsection{Unified Accounting Rules}
Let $n$ denote the sequence length (in tokens). For token-wise mechanisms, let $D_{\max}$ denote an upper bound on the per-token Euclidean distance \emph{in the metric space used by the corresponding baseline} (computed per dataset). We enforce a global budget constraint $B$ (e.g., $B=150$) and derive operational parameters as follows:

\paragraph{$d_{\chi}$-privacy (Sequential Baseline).}
Following prior work, we treat an entire prompt as one record (record-level adjacency) and privatize it token-wise.
Under sequential composition across $n$ token mechanisms, the worst-case privacy loss scales linearly with $n$.
To satisfy the global budget $B$, the per-token privacy parameter must be scaled down:
\begin{equation}
    \epsilon_{\text{token}} = \frac{B}{n \cdot D_{\max}}.
\end{equation}
For long sequences (e.g., $n=200$), this results in a small $\epsilon_{\text{token}}$, forcing excessive noise that destroys utility (the linear growth problem)~\cite{zeng2025privacyrestore}.

\paragraph{PrivacyRestore (Constant Baseline).}
Following~\citet{zeng2025privacyrestore}, PrivacyRestore aggregates sensitive information into a fixed-size meta-vector, so the protected unit is a single vector independent of $n$. We $\ell_2$-normalize the meta-vector before perturbation, so for any two adjacent meta-vectors $u,u'$, $\|u-u'\|_2 \le 2$. For vector mechanisms on $\ell_2$-normalized embeddings, enforcing a worst-case log-loss target $B$ implies:
\begin{equation}
    2\epsilon_{\mathrm{PR}} \le B
    \quad\Rightarrow\quad
    \epsilon_{\mathrm{PR}} = \frac{B}{2}.
\end{equation}

\paragraph{PPFT (Ours: Slot-wise Metric-DP with Per-vector Calibration).}
PPFT privatizes the pooled embedding interface produced by a client-side encoder.
Let $X$ be the input text and let $\mathbf{H}=\mathrm{Enc}(X)\in\mathbb{R}^{n\times d_e}$ be contextual token embeddings.
We apply non-overlapping $k$-pooling to obtain $m=\lceil n/k\rceil$ slot vectors
$\mathbf{U}=[\mathbf{u}_1,\dots,\mathbf{u}_m]$.

\textbf{Noise injection (matches the main text).}
For each row vector $\mathbf{u}_j$, we add isotropic $\ell_2$-Laplace noise by sampling a direction uniformly from the unit sphere
and a magnitude from a Gamma distribution (shape $d_e$, rate $\epsilon$), and then apply $\ell_2$ re-normalization as post-processing:
\begin{equation}
\begin{aligned}
\tilde{\mathbf{u}}_j \;=\; \mathrm{Renorm}\bigl(\mathbf{u}_j + \mathbf{N}_j\bigr),\qquad  \\
\mathbf{N}_j \sim \mathrm{Laplace}_{\ell_2}(\epsilon). 
\end{aligned}
\end{equation}

\textbf{Propagation across slots.}
Because $\mathrm{Enc}(\cdot)$ is contextual, a one-token substitution in $X$ can perturb many token embeddings, and consequently multiple pooled slots may change.
Therefore, PPFT does not assume that only one slot differs.
Instead, in Appendix~\ref{app:proof} we show that each slot mechanism satisfies metric-DP and that the
log-loss composes additively over the number of affected slots: if at most $s$ slots differ, the worst-case log-loss is bounded by $2\epsilon s$
under unit-norm boundedness.

\textbf{Budget alignment.}
For comparison with constant-size vector baselines (PrivacyRestore), we calibrate PPFT to match a \emph{per-vector} worst-case log-loss target $B$.
Under $\ell_2$-bounded slot vectors (e.g., unit-norm clipping/normalization in the transmission space),
$\|\mathbf{u}_j-\mathbf{u}'_j\|_2\le 2$ implies that a single released vector incurs worst-case log-loss at most $2\epsilon$.
Thus, enforcing the global target $B$ per exposed vector yields:
\begin{equation}
2\epsilon_{\mathrm{PPFT}} \le B
\quad\Rightarrow\quad
\epsilon_{\mathrm{PPFT}}=\frac{B}{2}=75.0.
\end{equation}
We empirically validate that this setting sufficiently resists inversion attacks in Section~\ref{sec:privacy_results}.

\subsection{Interpretation of $\epsilon$ in Embedding-space Metric DP}
Note that $\epsilon$ values are not directly comparable across DP instantiations with different metrics, normalizations, and units.
In high-dimensional embedding spaces, small $\epsilon$ can induce noise whose norm overwhelms semantic signal, causing severe utility collapse.
Prior work on metric DP for text representations commonly operates in higher-$\epsilon$ regimes to retain utility while preserving indistinguishability among nearby points in the embedding metric~\cite{feyisetan2020privacy}.
Empirically, in our inversion-attack evaluation (Section~4.4), reconstruction remains low (ROUGE-L $<0.25$) even at $\epsilon=75$.

See Appendix~\ref{app:proof} for the formal derivations.

\section{Privacy Accounting and Hyperparameters}
\label{app:privacy_hparams}
\begin{table}[t]
\centering
\small
\setlength{\tabcolsep}{5pt}
\renewcommand{\arraystretch}{1.2}
\resizebox{\linewidth}{!}{%
\begin{tabular}{lcccc}
\toprule
\textbf{Dataset} & $n$ & $D_{\max}$ & $\epsilon_{d_{\chi}}\!=\!\frac{150}{n D_{\text{max}}}$ & $\tau\!=\!\frac{2n}{150}$  \\
\midrule
Pri-DDXP & 106.00 & 1.64 & 0.863 & 1.413  \\
Pri-NLICE   & 72.00  & 1.39 & 1.499 & 0.960  \\
Pri-SLJA    & 193.00 & 1.45 & 0.536 & 2.573 \\
SQuAD       & 178.78 & 1.70 & 0.494 & 2.384  \\
CSQA        & 48.43  & 1.68 & 1.844 & 0.646  \\
\bottomrule
\end{tabular}
}
\caption{Dataset-specific hyperparameters aligned to budget $B=150$.
$n$: max token length used for accounting.
$D_{\max}$: an upper bound on per-token embedding distance in the metric space used by the $d_{\chi}$ baseline.
$\epsilon_{d_{\chi}}$ and $\tau$ are adjusted per dataset to maintain fixed $B$.}
\label{tab:privacy_hparams}
\end{table}

We align all methods to the same target budget $B=150$.
Table~\ref{tab:privacy_hparams} summarizes the dataset-specific statistics ($n$, $D_{\max}$) and the resulting hyperparameters derived below.

\paragraph{$d_{\chi}$-privacy (Full Text).}
Using the sequential composition bound over $n$ token mechanisms, we solve
$n \cdot \epsilon_{\text{token}} \cdot D_{\max} = B$ to find:
\begin{equation}
  \epsilon_{d_{\chi}} \;=\; \epsilon_{\text{token}} \;=\; \frac{B}{n \cdot D_{\max}}.
\end{equation}

\paragraph{Paraphrase.}
Using the proxy rule $2n/\tau = B$, we set the temperature as:
\begin{equation}
  \tau \;=\; \frac{2n}{B}.
\end{equation}

\paragraph{PrivacyRestore \& PPFT.}
PrivacyRestore releases a single fixed-size meta-vector, so the accounting is independent of $n$.
After $\ell_2$ normalization, $\|u-u'\|_2 \le 2$ implies a worst-case log-loss bound of at most $2\epsilon$.

PPFT releases a sequence of obfuscated slot vectors $\tilde{\mathbf{U}}=[\tilde{\mathbf{u}}_1,\dots,\tilde{\mathbf{u}}_m]$ by adding isotropic $\ell_2$-Laplace noise to each slot and applying $\ell_2$ re-normalization as post-processing.
Each slot mechanism admits a metric-DP bound (Appendix~\ref{app:proof}), and if at most $s$ slots differ, the worst-case log-loss scales as $2\epsilon s$ under unit-norm boundedness.
For numerical alignment with constant-vector baselines, we calibrate PPFT to the same \emph{per-vector} target $B$:
\begin{equation}
  \epsilon_{\text{PR}} \;=\; \epsilon_{\text{PPFT}} \;=\; \frac{B}{2} \;=\; 75.00.
\end{equation}

\section{Theoretical Analysis of PPFT under $\ell_2$-Laplace Noise}
\label{app:proof}

We analyze PPFT under the exact noise injection procedure described in the main text:
slot-wise isotropic $\ell_2$-Laplace noise followed by $\ell_2$ re-normalization as post-processing.

\subsection{Mechanism Definition}
Let $X$ be an input text and $\mathbf{H}=\mathrm{Enc}(X)\in\mathbb{R}^{n\times d_e}$ contextual token embeddings.
Non-overlapping $k$-pooling yields $m=\lceil n/k\rceil$ slot vectors $\mathbf{U}=[\mathbf{u}_1,\dots,\mathbf{u}_m]$.

For each slot, we sample isotropic $\ell_2$-Laplace noise by drawing a direction uniformly on the unit sphere
and a radius from a Gamma distribution (shape $d_e$, rate $\epsilon$), which is equivalent to the density form
$p(\mathbf{n})\propto \exp(-\epsilon\|\mathbf{n}\|_2)$.
We then output the obfuscated embedding via post-processing renormalization:
\begin{equation}
\begin{aligned}
\mathbf{y}_j &= \mathbf{u}_j + \mathbf{N}_j, \\
p(\mathbf{y}_j\mid \mathbf{u}_j)
&\propto \exp\!\left(-\epsilon\|\mathbf{y}_j-\mathbf{u}_j\|_2\right), \\
\tilde{\mathbf{u}}_j &= \frac{\mathbf{y}_j}{\|\mathbf{y}_j\|_2}.
\end{aligned}
\label{eq:ppft_mech}
\end{equation}

The full output is $\tilde{\mathbf{U}}=[\tilde{\mathbf{u}}_1,\dots,\tilde{\mathbf{u}}_m]$, and slots are perturbed independently.

\subsection{Per-slot Metric-DP Guarantee and Composition}
\paragraph{Per-slot metric-DP}
For any two slot vectors $\mathbf{u},\mathbf{u}'$ and any measurable set $\mathcal{S}$,
the pre-normalization mechanism in Eq.~\eqref{eq:ppft_mech} satisfies metric DP:
\begin{equation}
\begin{aligned}
P(\mathbf{y}\in\mathcal{S}\mid \mathbf{u}) \\
\le \exp\!\left(\epsilon\|\mathbf{u}-\mathbf{u}'\|_2\right)\, 
P(\mathbf{y}\in\mathcal{S}\mid \mathbf{u}').
\end{aligned}
\end{equation}

\begin{proof}
Using $p(\mathbf{y}\mid \mathbf{u})\propto \exp(-\epsilon\|\mathbf{y}-\mathbf{u}\|_2)$,
\[
\ln\frac{p(\mathbf{y}\mid \mathbf{u})}{p(\mathbf{y}\mid \mathbf{u}')}
=\epsilon\bigl(\|\mathbf{y}-\mathbf{u}'\|_2-\|\mathbf{y}-\mathbf{u}\|_2\bigr)
\le \epsilon\|\mathbf{u}-\mathbf{u}'\|_2,
\]
where the inequality follows from the reverse triangle inequality.
\end{proof}

\paragraph{Post-processing.}
The renormalization $\tilde{\mathbf{u}}=\mathbf{y}/\|\mathbf{y}\|_2$ is deterministic post-processing, so it does not weaken the above metric-DP guarantee.

\paragraph{Slot-sequence composition bound}
Because slots are perturbed independently, for two sequences $\mathbf{U},\mathbf{U}'$ we have:
\begin{equation}
\ln\frac{P(\tilde{\mathbf{U}}\mid \mathbf{U})}{P(\tilde{\mathbf{U}}\mid \mathbf{U}')}
\le \epsilon \sum_{j=1}^{m}\|\mathbf{u}_j-\mathbf{u}'_j\|_2.
\end{equation}
If at most $s$ slots differ and each slot vector is $\ell_2$-bounded so that $\|\mathbf{u}_j-\mathbf{u}'_j\|_2\le 2$,
then the worst-case log-loss is bounded by $2\epsilon s$.

\paragraph{Implication for budget alignment.}
In practice, a one-token substitution can affect multiple slots due to contextual encoding, so $s$ may exceed 1.
In our budget alignment (Appendix~\ref{app:privacy_budget}), we match a per-vector worst-case log-loss target $B$
(i.e., $2\epsilon \le B$) to ensure numerical comparability with constant-vector baselines, and empirically validate inversion resistance.

\section{Inverse Attack}
\label{app:Inverse Attack}

\paragraph{Threat model.}
Following prior work on embedding inversion~\cite{morris2023text, li2023sentence},
we consider an attacker who observes the representation transmitted by the client
(e.g., an embedding, an obfuscated query, or an auxiliary vector) and attempts to reconstruct
the user prompt (including privacy-sensitive content) using a generative model.
Concretely, we instantiate the attacker as \href{https://huggingface.co/openai-community/gpt2-medium}{\texttt{openai-community/gpt2-medium}}, a GPT-2 model~\cite{radford2019language}, which is fine-tuned to generate the original text from the observed signal.

\paragraph{Common attacker configuration.}
Across all methods, we use \textbf{GPT2-medium} as the attack model,
trained for \textbf{20 epochs} with learning rate \textbf{1e-5} and batch size \textbf{32}.
During generation, we use greedy decoding with maximum generation length \textbf{256}.
The attacker is trained on the corresponding training split and evaluated on the test split.

\begin{table*}[t]
\centering
\small
\setlength{\tabcolsep}{4.5pt}
\begin{tabular}{lcccccccc}
\toprule
Pooling size & \(\epsilon{=}0.01\) & \(\epsilon{=}0.46\) & \(\epsilon{=}0.86\) & \(\epsilon{=}2.01\) & \(\epsilon{=}2.29\) & \(\epsilon{=}17.2\) & \(\epsilon{=}22.93\) & \(\epsilon{=}75.0\) \\
\midrule
4  & 0.02974 & 0.03045 & 0.03178 & 0.03487 & 0.03373 & 0.16013 & 0.24380 & \textbf{0.43974} \\
8  & 0.05506 & 0.05554 & 0.05525 & 0.05920 & 0.06266 & 0.09974 & 0.15784 & 0.33750 \\
16 & 0.05039 & 0.05177 & 0.04938 & 0.04910 & 0.05055 & 0.14032 & 0.15935 & 0.17990 \\
\bottomrule
\end{tabular}
\caption{Noise-aware inverse attack results (ROUGE-L). The attacker is trained with noisy representations while we report reconstruction quality under different privacy budgets at inference.}
\label{tab:noise_aware_inverse}
\end{table*}

\begin{table*}[t]
\centering
\small
\setlength{\tabcolsep}{6pt}
\begin{tabularx}{\textwidth}{c X X}
\toprule
\textbf{Ex.} & \textbf{Ground truth} & \textbf{Reconstruction (blue=same, red=different)} \\
\midrule

1 &
A 46-year-old male has a history of chronic pancreatitis, diabetes, obesity, pancreatic cancer in family members.
The 46-year-old male presents the symptoms of cough, diarrhea, nausea, pain, pale stools and dark urine, skin lesions, underweight.
What is the likely diagnosis?
&
\same{A }\diff{6}\same{-year-old }\diff{woman }\same{has a history of }\diff{smoking, }\same{diabetes, }\diff{high blood pressure, }\same{obesity, }\diff{high cholesterol, high blood pressure, smoking. }
\same{The }\diff{6}\same{-year-old }\diff{woman }\same{presents the symptoms of }\same{cough, }\diff{fever, fatigue, }\same{pain, }\diff{shortness of breath, }\same{skin lesions. }
\same{What is the likely diagnosis?}
\\

\midrule

2 &
A 45-year-old woman has a history of chronic pancreatitis, diabetes, obesity, pancreatic cancer in family members, smoking.
The 45-year-old woman presents the symptoms of diarrhea, fatigue, nausea, pain, pale stools and dark urine, skin lesions, underweight.
What is the likely diagnosis?
&
\same{A }\diff{22}\same{-year-old }\diff{man }\same{has a history of }\diff{alcohol addiction, }\same{smoking, }\diff{alcohol addiction, heart failure, heart valve issue. }
\same{The }\diff{22}\same{-year-old }\diff{man }\same{presents the symptoms of }\diff{chest }\same{pain, }\diff{shortness of breath, }\same{pain, fatigue, }\diff{shortness of breath with exertion, }\diff{\ldots}
\\

\bottomrule
\end{tabularx}
\caption{Qualitative examples for the noise-aware inverse attack.
\textcolor{blue}{Blue} indicates spans that exactly match the original prompt,
whereas \textcolor{red}{red} indicates mismatched or hallucinated content, including medically salient details.}
\label{tab:inverse_attack_examples_colored}
\end{table*}

\paragraph{Attack on PPFT (ours).}
For PPFT, the attacker operates on the same noisy, pooled embedding representation that is exposed to the server.
Specifically, we reuse the encoder and $k$-pooling module from the Stage~1-aligned LLaMA-1B PPFT model to process the input, producing pooled encoder representations identical to those used by PPFT.
These pooled embeddings are then passed through a learnable projection layer that maps them to the input embedding space of GPT2-medium, which serves as the attacker decoder.
During attack training, the encoder is kept frozen, while only the projection layer and GPT2-medium are optimized.
The attacker is trained end-to-end to perform sequence reconstruction, learning to generate the original prompt text from the observed noisy and pooled embeddings.

\paragraph{Attack on PrivacyRestore.}\mbox{}\\
PrivacyRestore~\cite{zeng2025privacyrestore} transmits an incomplete user query in which privacy-sensitive spans are removed, together with a \emph{meta vector} that encodes information about the removed spans.
To match the inference-time observable interface of PrivacyRestore, our inversion attacker is conditioned on \emph{both} the incomplete query and the corresponding meta vector, and is trained to reconstruct the original full query. Specifically, we encode the masked query with the attacker decoder in the standard autoregressive manner, while a learnable projection layer maps the meta vector to the hidden-state dimension of GPT2-medium and injects it as an auxiliary conditioning signal. We jointly fine-tune GPT2-medium and the projection layer under the common attacker configuration to generate the original prompt text from the observable pair.

\paragraph{Attack on $d_\chi$-privacy and Paraphrase.}
For $d_\chi$-privacy, the client transmits an obfuscated text query obtained by applying token-level privatization, where each token is replaced by a randomized alternative sampled according to a distance-based distribution in an embedding space~\cite{feyisetan2020privacy}.
For Paraphrase, the client transmits a paraphrased version of the original query generated by a client-side model.
In both cases, the attacker observes only text and directly uses the garbled or paraphrased query as input context to GPT2-medium, which is then fine-tuned to reconstruct the original prompt text using the same attack training procedure described above.

\paragraph{Evaluation metric.}
We quantify inversion effectiveness using ROUGE-L as a sequence-level reconstruction metric, measuring similarity between the attack model's generated output and the ground-truth original prompt on the test split.
Higher ROUGE-L indicates more successful surface-level reconstruction and thus weaker privacy protection.
Attribute-level reconstruction metrics are reported separately to assess the recovery of specific sensitive information.

\begin{table*}[t]
\centering
\small
\setlength{\tabcolsep}{8pt}
\renewcommand{\arraystretch}{1.2}
\begin{tabularx}{\textwidth}{c X X}
\toprule
\textbf{Ex.} & \textbf{Ground truth} & \textbf{Reconstruction (blue=same, red=different)} \\
\midrule

1 &
A 57-year-old male has a history of antipsychotic medication usage, nausea, stimulant drug use.
The 57-year-old male presents the symptoms of involuntary eye movement, jaw pain, muscle spasms, muscle spasms in neck, ptosis, shortness of breath.
What is the likely diagnosis?
&
\diff{The diagnosis of the }\same{57-year-old male}\diff{ who has been experiencing symptoms of }
\diff{eye jumping, unknown button, joint pain and }\same{muscle spasms in neck}\diff{, is psychosis. }
\same{What is the }\diff{diagnosis?}
\\

\midrule

2 &
A 8-year-old woman has a history of active cancer, deep vein thrombosis, hormone intake, immobility for $>$3 days, surgery within last month.
The 8-year-old woman presents the symptoms of coughing up blood, loss of consciousness, pain, shortness of breath, swelling.
What is the likely diagnosis?
&
\diff{The patient has been in the hospital for over 3 weeks, with intravenous drug use, migraine, intake of bed, }\same{surgery}\diff{. }
\diff{The patient's symptoms are cough, fever, }\same{pain}\diff{, }\same{swelling}\diff{. }
\same{What is the likely diagnosis?}
\\

\bottomrule
\end{tabularx}
\caption{Qualitative examples for the Stage-1 aligned inversion attacker.
\textcolor{blue}{Blue} spans exactly match the original prompt, while \textcolor{red}{red} spans differ.
Even with a stronger attacker aligned to the encoder space, reconstructions often preserve only partial lexical overlaps rather than medically faithful recovery.}
\label{tab:inverse_stage1_examples}
\end{table*}

\section{Noise-Aware Inverse Attack Training}
\label{app:inverse_noise_aware}

In this additional experiment, we strengthen the adversary by allowing it to train the inverse attack model on noisy representations.
All experiments are conducted on the Pri-DDX dataset.
Concretely, we keep the inverse model architecture and training procedure identical to the main inverse-attack setting in Appendix~\ref{app:Inverse Attack},
but inject the same privacy noise during attacker training (i.e., the attacker is trained with representations perturbed under \(\epsilon = 75\)).
This setting tests whether a noise-aware attacker—one that has access to the defense mechanism and can adapt to it—can substantially improve
reconstruction of the original input text.

\paragraph{Quantitative results.}
Table~\ref{tab:noise_aware_inverse} reports ROUGE-L reconstruction scores as a sequence-level similarity metric across privacy budgets and pooling sizes.
Overall, the noise-aware attacker achieves higher ROUGE-L than a noise-unaware attacker, especially in the weak-noise regime (large \(\epsilon\)).
However, even with noise-aware training, the attacker does not recover the full original text: performance remains low for strong noise (small \(\epsilon\)),
and improvements at the inference-time privacy setting (\(\epsilon=75\)) remain far from exact reconstruction.
Among pooling strategies, pooling-4 is the most vulnerable (\(0.4397\) at \(\epsilon=75\)),
pooling-8 is intermediate (\(0.3375\)),
and pooling-16 is the most robust (\(0.1799\)).
This trend is consistent with the intuition that larger pooling sizes induce stronger information compression,
making exact inversion intrinsically harder even when the attacker matches the training-time noise distribution.

Importantly, we also conducted a matched noise-aware comparison for PrivacyRestore under the same inference-time privacy setting (\(\epsilon=75\)).
Under this stronger attacker, PrivacyRestore reaches a substantially higher reconstruction score (ROUGE-L up to \(0.72\)),
whereas PPFT remains markedly lower across all pooling settings.
This comparison is critical because it shows that the stronger attack does not simply increase reconstruction for all methods uniformly; rather, PPFT retains a clear advantage even when the adversary is fully aware of the defense mechanism and trained on noise-corrupted representations.

These results also highlight an important caveat: ROUGE-L can be inflated when the attacker learns to replicate common scaffolding tokens and templates,
even if the recovered content is factually inconsistent with the original private text.
Therefore, while noise-aware training increases lexical overlap, it does not imply faithful reconstruction.
Taken together with the matched PrivacyRestore comparison, our results show that PPFT provides substantially stronger reconstruction resistance under realistic privacy-preserving inference conditions.

\paragraph{Qualitative analysis: template-matching rather than true recovery.}
Despite higher ROUGE-L at large \(\epsilon\), outputs often improve by mimicking the \emph{surface form} of the data (e.g., age/gender template and symptom-list scaffolding),
rather than recovering correct patient attributes or medical history.
Table~\ref{tab:inverse_attack_examples_colored} provides two representative cases, where tokens identical to the ground truth are highlighted in \textcolor{blue}{blue},
while mismatched or hallucinated content is highlighted in \textcolor{red}{red}.
As shown, the attacker frequently reproduces high-frequency structural phrases (e.g., ``has a history of'', ``presents the symptoms of'', and the question suffix),
yet changes medically salient details such as age, gender, comorbidities, and symptom composition.


\begin{figure}[t]
    \centering
    \includegraphics[width=1.0\linewidth]{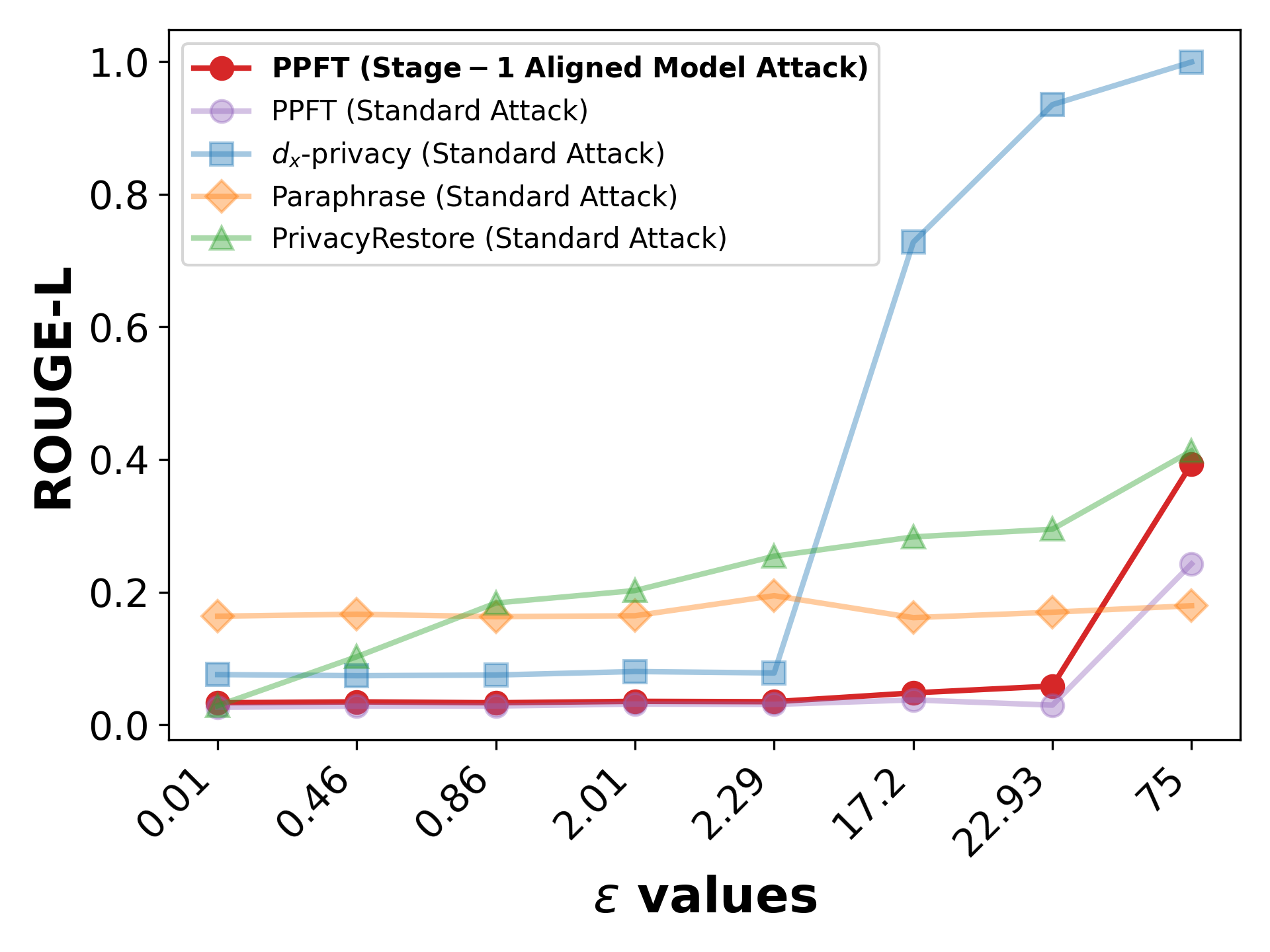}
    \caption{Inversion attacks on PPFT using a Stage-1 aligned model (stronger attacker) under varying privacy budgets $\epsilon$. For comparison, we also report inversion results from a GPT-2 Medium model (weaker attacker).}
    \label{fig:noise_var_stage1}
\end{figure}

\section{Inversion Attack with a Stage-1 Aligned Model}
\label{app:inverse_stage1_aligned}

In this additional setting, we consider a stronger adversary that better reflects a realistic threat model for LLM service providers.
Specifically, we assume the provider is willing to recover user prompts and thus replaces the inversion attacker (GPT-2 Medium in Appendix~\ref{app:Inverse Attack})
with a \emph{Stage-1 aligned model}---i.e., a decoder already aligned to the encoder representations during Stage~1.
This attacker starts from a substantially more favorable initialization since it has been explicitly trained to interpret the encoder-aligned latent space.
All other training and evaluation conditions follow Appendix~\ref{app:Inverse Attack}.

\paragraph{Quantitative results.}
Figure~\ref{fig:noise_var_stage1} reports ROUGE-L reconstruction scores across privacy budgets.
While the Stage-1 aligned attacker slightly improves reconstruction quality in the weak-noise regime,
it still fails to faithfully recover the original prompt.
Notably, under the inference-time condition (\(\epsilon=75.0\)), ROUGE-L reaches \(0.393\), remaining below \(0.4\).

\paragraph{Qualitative analysis.}
Table~\ref{tab:inverse_stage1_examples} shows representative reconstructions.
Spans that \emph{exactly match} the original prompt are highlighted in \textcolor{blue}{blue},
whereas altered or hallucinated content is highlighted in \textcolor{red}{red}.
Even with the Stage-1 aligned attacker, improvements in ROUGE-L largely come from reproducing a subset of frequent tokens or local phrases,
while medically salient attributes (e.g., history and symptom composition) are not reliably recovered.

\section{Universal Zero-shot Embedding Inversion under Token Pooling}
\label{app:pooling_inversion}

Recent work has shown that text embeddings can be inverted to recover substantial semantic information about the original inputs, even under black-box access assumptions \cite{morris2023text, zhang2025universal}. 
These attacks, however, are primarily studied under encoders that map an entire input sequence to a \emph{single} embedding vector.
In this appendix, we examine whether such inversion techniques remain effective when the encoder employs \emph{token pooling}, producing multiple embeddings per input.

\paragraph{Threat Model.}
We consider a black-box adversary who has access to (i) the pooled embeddings of a private input and (ii) query access to the same encoder used to generate those embeddings.
This setting is consistent with prior embedding inversion work \cite{morris2023text, zhang2025universal}, but differs in that the encoder applies pooling over fixed-size token blocks ($k{=}4$ in our experiments), followed by noise injection.
The adversary attempts to reconstruct the original text using iterative, embedding-guided decoding.

\paragraph{Experimental Setup.}
We conduct two inversion experiments on the Pri-NLICE dataset introduced by \citet{zeng2025privacyrestore}.
In both cases, the target encoder is a LoRA-adapted \texttt{Llama-3.2-1B-Instruct} model with pooling size $k{=}4$ and Laplace noise injection ($\epsilon{=}75$).
For generation, we use \href{https://huggingface.co/meta-llama/Llama-3.2-3B-Instruct}{\texttt{meta-llama/Llama-3.2-3B-Instruct}} as the decoder.
To ensure a fair comparison, we use the same privacy parameter $\epsilon$ for inversion experiments as in the inference setting reported in Table~\ref{tab:main_results}.

Following the adversarial decoding paradigm of \citet{zhang2025universal}, we perform iterative inversion for up to 10 iterations.
At each iteration, the decoder generates candidate texts using embedding-guided search, and the highest-scoring candidate (based on cosine similarity in embedding space) is selected and used as the seed for the next iteration.
Reconstruction quality is evaluated using ROUGE-L against the ground-truth text, averaged over the dataset. 
\begin{figure}
    \centering
    \includegraphics[width=1.0\linewidth]{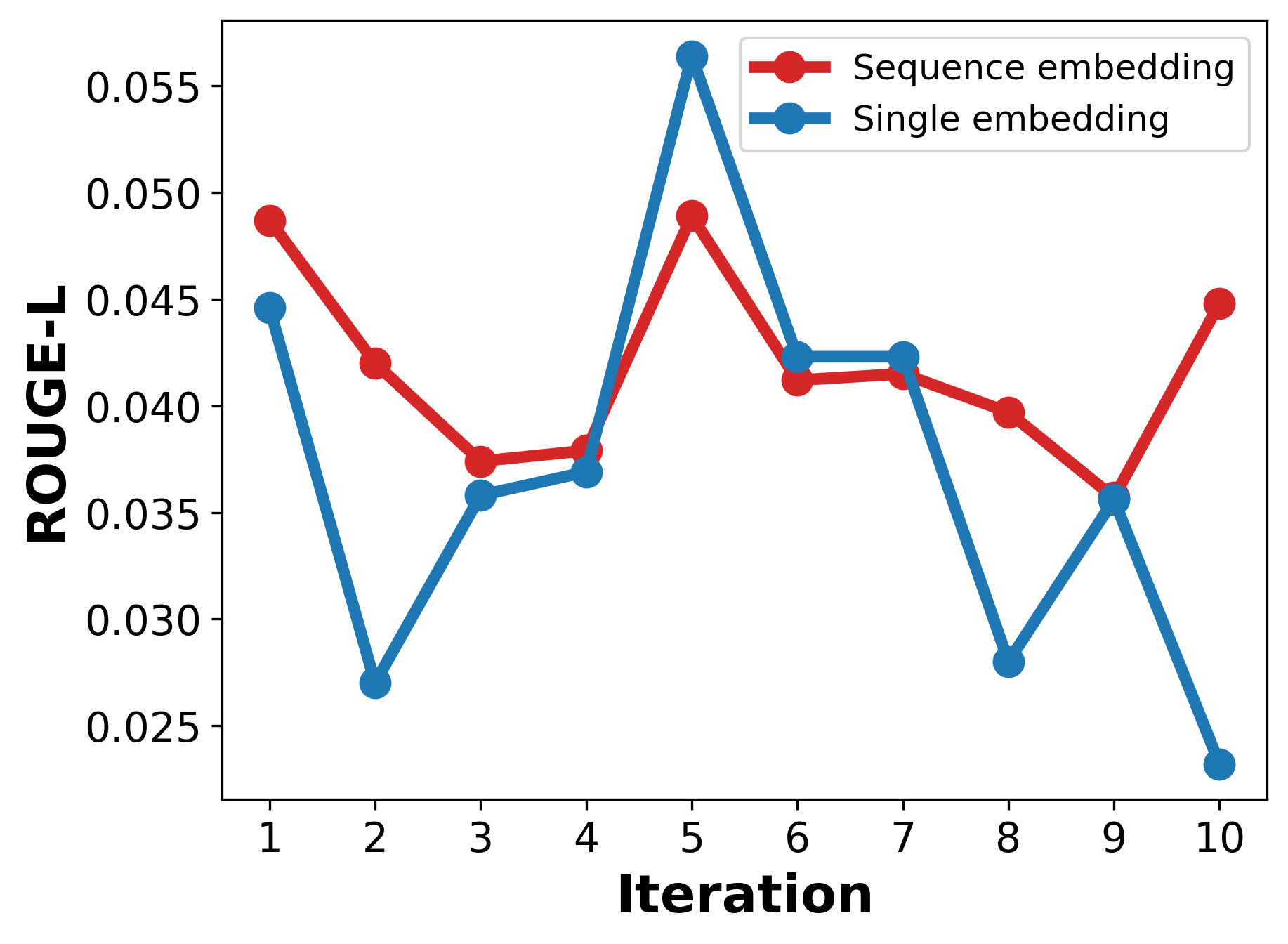}
    \caption{Reconstruction quality across iterations for the pooled-embedding (Experiment~1) and single-embedding (Experiment~2) settings.}
    \label{fig:iter_ROUGE}
\end{figure}
\paragraph{Experiment 1: Pooling-Aligned Inversion.}
In the first experiment, we directly attack the pooled representation.
The encoder outputs a \emph{sequence of pooled embeddings} (one per 4 tokens), and during inversion we compute cosine similarity block-wise between generated and target embeddings, aggregating scores across aligned blocks.
Generation is constrained to the original input length, ensuring that the number of pooled embeddings in the generated text does not exceed that of the target. Figure~\ref{fig:iter_ROUGE} reports ROUGE-L scores across iterations.



Despite iterative refinement, reconstruction quality remains low and does not exhibit a consistent upward trend.
This contrasts sharply with prior results on non-pooled encoders, where repeated iterations significantly improve lexical overlap \cite{zhang2025universal}.

\paragraph{Experiment 2: Mean-Pooled Single-Vector Inversion.}
To more closely match the setting of prior work, we perform a second experiment in which the pooled embeddings are averaged into a single vector after noise injection.
This removes the structural mismatch between pooled encoders and single-vector inversion methods.
Since the target representation is now a single embedding, we allow the decoder to generate up to 250 tokens, mirroring the unconstrained generation setting used in \citet{zhang2025universal}. Figure~\ref{fig:iter_ROUGE} reports ROUGE-L scores across iterations.



Although this setting removes the pooling mismatch, inversion performance remains poor.
Even at its peak (iteration 5), ROUGE-L remains below 0.06, and later iterations often degrade reconstruction quality.

\paragraph{Discussion.}
Across both experiments, embedding inversion fails to recover meaningful lexical information from pooled, noise-injected embeddings.
This is notable because the second experiment explicitly aligns with the assumptions of prior inversion attacks by collapsing the pooled representation into a single embedding.
The results suggest that the combination of token pooling and noise injection substantially alters the embedding landscape, making iterative, cosine-similarity-guided decoding ineffective.

From a security perspective, these findings indicate that pooling-based encoders provide a qualitatively stronger defense against embedding inversion than previously studied single-vector encoders.
In contrast to earlier conclusions that ``embeddings reveal (almost) as much as text'' \cite{morris2023text}, our results show that this claim does not directly extend to encoders that disrupt token-level alignment through pooling.

\end{document}